%% file: mnras_template.tex
\documentclass[fleqn,usenatbib]{mnras}

\usepackage{newtxtext,newtxmath}
\usepackage{chemformula}
\usepackage{longtable}
\usepackage[T1]{fontenc}

\DeclareRobustCommand{\VAN}[3]{#2}
\let\VANthebibliography\thebibliography
\def\thebibliography{\DeclareRobustCommand{\VAN}[3]{##3}\VANthebibliography}

\usepackage{graphicx}	
\usepackage{amsmath}	
\usepackage{color}
\usepackage{soul}
\usepackage{chemmacros}[2016/05/04]
\chemsetup{ modules = {reactions} }

\title[Biosignatures from TRAPPIST-1e]{Biosignatures from pre-oxygen photosynthesising life on TRAPPIST-1e}

\author[Jake K. Eager-Nash et al.]{Jake K. Eager-Nash,$^{1,2}$\thanks{E-mail: jeagernash@uvic.ca (JKE-N)}
Stuart J. Daines,$^{3}$
James W. McDermott,$^{1,7}$
Peter Andrews,$^{1}$
Lucy A. Grain,$^{1}$
\newauthor
James Bishop,$^{1}$
Aaron A. Rogers,$^{1}$
Jack W. G. Smith,$^{1}$
Chadiga Khalek,$^{4}$
Thomas J. Boxer,$^{4}$
\newauthor
Mei Ting Mak,$^{1}$
Robert J. Ridgway,$^{1}$
Eric H\'ebrard,$^{1}$
F. Hugo Lambert,$^{5}$
Timothy M. Lenton,$^{6}$
\newauthor
and
Nathan J. Mayne,$^{1}$
\\
$^{1}$Department of Physics and Astronomy, University of Exeter, Exeter, UK\\
$^{2}$School of Earth and Ocean Sciences, University of Victoria, Victoria, BC, Canada \\
$^{3}$ University of Exeter, Exeter, UK\\
$^{4}$Natural Sciences, University of Exeter, Exeter, UK\\
$^{5}$Department of Mathematics and Statistics, University of Exeter, Exeter, UK\\
$^{6}$Global Systems Institute, University of Exeter, Exeter, UK\\
$^{7}$The Natural History Museum, London, UK\\
}

\date{Accepted XXX. Received YYY; in original form ZZZ}

\pubyear{2023}

\begin{document}
\label{firstpage}
\pagerange{\pageref{firstpage}--\pageref{lastpage}}
\maketitle

\begin{abstract}
In order to assess observational evidence for potential atmospheric biosignatures on exoplanets, it will be essential to test whether spectral fingerprints from multiple gases can be explained by abiotic or biotic-only processes. Here, we develop and apply a coupled 1D atmosphere-ocean-ecosystem model to understand how primitive biospheres, which exploit abiotic sources of \ch{H2}, \ch{CO} and \ch{O2}, could influence the atmospheric composition of rocky terrestrial exoplanets. We apply this to the Earth at 3.8\,Ga and to TRAPPIST-1e. We focus on metabolisms that evolved before the evolution of oxygenic photosynthesis, which consume \ch{H2} and \ch{CO} and produce potentially detectable levels of \ch{CH4}. \ch{O2}-consuming metabolisms are also considered for TRAPPIST-1e, as abiotic \ch{O2} production is predicted on M-dwarf orbiting planets. We show that these biospheres can lead to high levels of surface \ch{O2} (approximately 1--5\,\%) as a result of \ch{CO} consumption, which could allow high \ch{O2} scenarios, by removing the main loss mechanisms of atomic oxygen. Increasing stratospheric temperatures, which increases atmospheric \ch{OH} can reduce the likelihood of such a state forming. \ch{O2}-consuming metabolisms could also lower \ch{O2} levels to around 10\,ppm and support a productive biosphere at low reductant inputs. Using predicted transmission spectral features from \ch{CH4}, \ch{CO}, \ch{O2}/\ch{O3} and \ch{CO2} across the hypothesis space for tectonic reductant input, we show that biotically-produced \ch{CH4} may only be detectable at high reductant inputs. \ch{CO} is also likely to be a dominant feature in transmission spectra for planets orbiting M-dwarfs, which could reduce the confidence in any potential biosignature observations linked to these biospheres.

\end{abstract}


\section{Introduction}

Planets orbiting M-dwarfs provide the most realistic opportunity for detecting biosignatures in the near future, through JWST and other telescopes in development, such as the European Extremely Large Telescope (ELT). JWST has already been used to determine that TRAPPIST-1b+c are unlikely to have thick atmospheres \citep{Greene2023,Zieba2023}, but could have surface pressures of up to 10\,bar \citep{Ih2023,Lincowski2023}. In order to detect potential life on these planets, which are orbiting very different host stars to our Sun, it is important to not only study the potential abiotic atmospheres of these planets, but also the potential relationship between life, the atmosphere and the host star \citep{Catling2018}.

Life on other planets may most likely resemble life on the early Earth, rather than the modern day. Life was thought to have emerged during the early Archean, by 3.7\,Ga at the latest \citep{Rosing1999}, with potential for life to have emerged much earlier \citep{Knoll2017}. Since then, life on Earth has diversified hugely, undergoing major evolutionary revolutions including the evolution of oxygenic photosynthesis and multi-cellular eukaryotes \citep{LentonWatson2011}. The number of evolutionary steps, as well as their potential difficulty, which are required to get to higher levels of complexity means that many planets may be limited to more primitive life \citep{Watson2008}. Simple microbial biospheres similar to those that existed during the Archean could be the most common.

The detection of primitive biospheres via their effect on \ch{CH4} and \ch{CO} has been considered for planets (around Sun-like stars), with ecosystems driven by \ch{H2} and \ch{CO} consumption, leading to atmospheres with a low \ch{CO}:\ch{CH4} ratio \citep{Sauterey2020,Thompson2022}. \citet{Krissansen-Totton2018earth-biosig} predicts that biospheres could produce a detectable disequilibrium with an atmosphere composed of \ch{CO2} and \ch{CH4} in the absence of \ch{CO}. This may be the most likely source of biosignature detection with current instrumentation \citep{Krissansen-Totton2018t1e-biosig}. A potential additional challenge for the unambiguous identification of biosignatures specific to planets around M-dwarf stars is the accumulation of abiotic \ch{CO} and possibly \ch{O2}. This results from a higher photolysis of \ch{CO2} by a higher FUV flux from the M-dwarf spectrum, combined with a lower rate of recombination of these species due to a lower abundance of \ch{OH} resulting from the lower NUV flux emitted by M-dwarfs \citep{Harman2015}. \citet{Schwieterman2019} finds that high levels of \ch{CO} could accumulate on inhabited M--dwarf orbiting planets. Using General circulation models (GCMs) and photochemical models, \citet{Fauchez2019} found that clouds and especially hazes could play an important role in blocking the observations of Archean-like atmospheres in large regions of JWST transmission spectra, however \ch{CO2} features are likely to be readily observable on such planets. 

A fundamental challenge to identifying a biosignature from atmospheric concentrations of \ch{CH4}, \ch{CO} and \ch{O2} is that all of these gases have both abiotic and biological sources and sinks, and may participate in biogeochemical cycles through the atmosphere-ocean-surface system. Their concentrations are therefore controlled by the combination of regulatory feedbacks generated by a combination of abiotic and hypothetical biological processes. As argued by \citet{Catling2018}, a consistent approach then requires populating a hypothesis space with predictions from exo-Earth system models, applying observational constraints and making full use of available context. This type of model has been applied to the early Earth around the early Sun \citep{Kharecha2005, Sauterey2020}, however there has yet to be an ecosystem model focusing on the implications of M-dwarfs spectra. 

Here, we use a newly developed coupled atmosphere-ocean-ecosystem model to show that a \ch{H2} and \ch{CO} consuming biosphere, which produces \ch{CH4} as a byproduct, leads to an atmosphere with detectable levels of \ch{CO2} and \ch{CH4}, but also large signals of \ch{CO} for a TRAPPIST-1e analogue. We review the abiotic and ecosystem processes we consider in Sections~\ref{sec:abioticprocesses} and \ref{sec:ecosystemprocesses}. In Section~\ref{sec:methods}, we outline the model components used here, with results from the coupled model and subsequent transmission spectra shown in Section~\ref{sec:results}. In Section~\ref{sec:discussion} we discuss the implications of large \ch{CO} features in our biotic configuration that can lead to the ambiguous detection of a potential biosphere.

\section{Abiotic background assumptions and processes}
\label{sec:abioticprocesses}

We consider a terrestrial planet in the habitable zone of an M-dwarf star, with an Earth-like atmosphere dominated by \ch{N2} and \ch{CO2}, with liquid water, noting that this may imply an atypical evolutionary history given the star's extended pre-main sequence phase \citep{LugerBarnes2015}. This most likely also requires a functioning recycling of the lithosphere of some kind, in order to cycle \ch{CO2} and generate a silicate weathering feedback.

The atmosphere's total hydrogen content will then be controlled by redox balance, i.e. the balance between net surface reductant input and hydrogen escape, and atmospheric composition by the combination of photochemistry and surface processes.

Outgassing at the planetary surface is uncertain. Modern Earth plate tectonics allow for the for efficient cycling of volatiles, which is thought to be important for habitability by controlling \ch{CO2} over geological time scales \citep{Walker1981}. However, the tectonic history of the Earth remains uncertain \citep[e.g.][]{PalinSantosh2021,LourencoRozel2023} and modern plate tectonics may only have evolved as late as 0.85\,Ga \citep{Korenaga2013}. During the Archean, the Earth could have been in a stagnant lid state \citep{Solomatov1995}, where the lithosphere is a single lid and there is little surface motion. In these conditions outgassing rates are thought to be lower \citep{Guimond2021}. Arguments have also been made that plate tectonic could have been preceded by a plutonic squishy lid, a regime that has small strong plates that are separated by warm and weak regions created by plutonism \citep{LourencoRozel2023}, similar to Venus \citep{HarrisBedard2014,Davies2023}. Modelling of this regime suggests that outgassing rates could be up to a factor of two higher, assuming an extrusion efficiency of the lithosphere similar to today \citep{Lourenco2020}. 

The mix of gases outgassed is affected by the fugacity of the mantle  \citep{Kasting1993redox,Guimond2021}. A more reduced mantle will outgas more \ch{H2} and \ch{CO} and less \ch{H2O} and \ch{CO2}, while a higher mantle fugacity would lead to the opposite \citep{Guimond2021}. A stagnant lid planet may have outgassed significantly less than a planet with plate tectonics \citep{Guimond2021}. Additional tectonically-driven reduced gases may be produced by serpentinization at mid-ocean ridges  \citep[potentially a large source of \ch{H2} and possibly \ch{CH4}, with \ch{H2} dominating on modern Earth][]{McCollomBach2009}.

The atmospheres of planets orbiting M-dwarfs receive a different stellar spectra compared to Earth, which has an effect on climate \citep{Shields2013,Eager-Nash2020}, as well as the atmospheric composition \citep{Segura2005,Harman2015,Rugheimer2015,Rugheimer2018,Schwieterman2019,Kozakis2022}. M-dwarfs have a lower proportion of radiation in the ultraviolet range, particularly in the mid and near UV (200-400\,nm) \citep{Wunderlich2020}, which is important for photochemistry \citep{Harman2015}. On M-dwarfs, this allows biosignature gases such as \ch{CH4}  to be able to be maintained at higher concentration for a given \ch{CH4} outgassing flux due to a lack of \ch{OH} from water vapour photolysis \citep{Segura2005,Meadows2018pc,Schwieterman2019}. 

For lifeless planets, it has been found that \ch{O2} may accumulate to significant levels in the atmosphere via the photolysis of \ch{CO2}, with a slow catalysed recombination of \ch{O} and \ch{CO}, which is caused by a higher ratio of Far UV (FUV, 110--200\,nm) to mid and near UV (200--400\,nm) compared to the Sun \citep{Tian2014,Domagal-Goldman2014,Harman2015,Ranjan2023}. This is caused by the lower \ch{H2O} photolysis resulting from the significantly reduced NUV flux relative to the Sun, which is able to reach close to the surface to photolyse \ch{H2O} \citep{Harman2015}. The NUV cross sections for \ch{H2O} were found to have been underestimated however, with \citet{Ranjan2020} finding that these increased rates led to higher \ch{OH} levels, which catalyse the recombination of \ch{CO} and \ch{O} to reform \ch{CO2}, and removes the possibility of \ch{O2} false positives. \ch{NO} can be feasibly produced by lightning in the atmosphere and could eradicate the \ch{O2} false positive, by catalysing the recombination of \ch{CO} and \ch{O} \citep{Harman2018}. \citet{Hu2020} went on to show that the inclusion of \ch{NO_x} reservoir species, \ch{HO2NO2} and \ch{N2O5}, could lead to large abiotic sources of atmospheric \ch{O2}. \ch{HO2NO2} and \ch{N2O5} act as reservoirs as they are relatively stable and store \ch{NO} in a form that cannot catalyse \ch{CO2} formation, and would rainout into the oceans. However, it was recently found that this result was due to the model top height being too low (54\,km), which leads to erroneously high levels of \ch{O} production at the model top \citep{Ranjan2023}. Thus, it is now considered unlikely that an \ch{O2} false positive can be produced by \ch{CO2} photolysis \citep{Ranjan2023}. 

Although M-dwarfs emit significantly less UV radiation than G-dwarfs, some of these stars can flare regularly \citep{Hawley2014}, which increases UV radiation incident on the planet and significantly impacts atmospheric composition \citep{Chen2020,Ridgway2022}. For a modern Earth-like atmospheric composition, flaring increases the formation of ozone in the stratosphere, which provides additional protection to the planet's surface from the UV radiation from subsequent flares \citep{Ridgway2022}. Taking a temporally averaged spectrum of a flaring M-dwarf produces a reasonable approximation of the mean chemical composition of an atmosphere experiencing flares \citep{Ridgway2022}.

\section{Ecosystem processes}
\label{sec:ecosystemprocesses}

Methanogens are thought to be one of the first organisms to have been present on Earth \citep{Battistuzzi2004}. Methanogens are \ch{CH4}-producing microscopic organisms and could have metabolised \ch{H2} from volcanic outgassing. \ch{CO}--consuming organisms are also thought to have evolved early in Earth's history \citep{Ferry2006,Lessner2006,Weiss2016}. These organisms had the potential to impact the composition of the atmosphere by providing a major source of \ch{CH4} \citep{Kharecha2005}, which impacts the climate \citep{Eager-Nash2023}. Models of plausible biospheres on early Earth, which include methanogens, have been used to investigate possible biological productivity and subsequent atmospheric conditions \citep{Kharecha2005,Ozaki2018,Sauterey2020}. \citet{Kharecha2005} found that early methanogenic biospheres could have converted the majority of lower atmospheric \ch{H2} to \ch{CH4}, producing atmospheres with \ch{CH4} concentrations from 10 to 3,500\,ppm. This is controlled by redox balance between reductant input and hydrogen escape, and with cycling of hydrogen through atmospheric \ch{CH4} photolysis and biosphere \ch{CH4} production. We assume that life on other planets may also utilise available \ch{H2} and \ch{CO} for metabolism, to understand the effect that these biospheres could have on planets that orbit M-dwarfs.

\begin{figure*}
    \centering
    \includegraphics[width=0.9\linewidth]{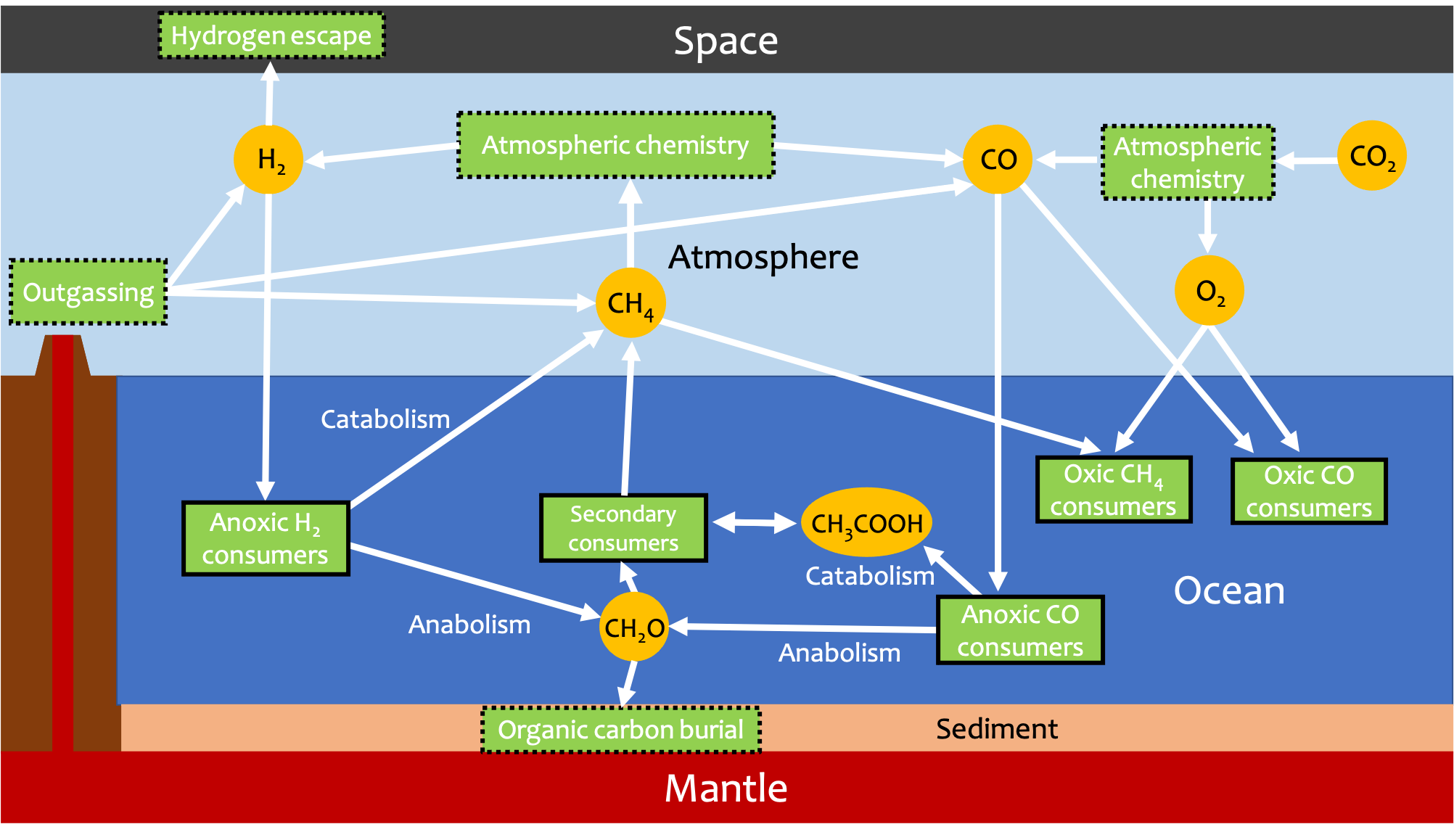}
    \caption{Schematic showing the biosphere reaction pathways and an overview of the interaction with the atmosphere captured in our modelling framework. Green boxes show processes, both biotic (dashed outline) and abiotic (solid outline), circles show reservoirs of species and arrows show fluxes between reservoirs via the different processes. Volcanic outgassing drives biospheric productivity by providing electron donors for primary producers. These are used for either catabolism to produce energy and \ch{CH4} as a waste product, with this energy used for biomass production, which is then either recycled by secondary consumers and eventually converted to \ch{CH4} again or the biomass is buried in the sediments.
    \label{fig:biosphere_schematic}
    }
\end{figure*}

Conceptually, a biosphere could function as summarised schematically in Figure~\ref{fig:biosphere_schematic}, which is the form we take in our model. We now describe a modelling framework that follows similar logic to \citet{Kharecha2005,Ozaki2018,Sauterey2020}, but is considered in more general terms to apply in an astrobiological context for pre-photosynthetic biospheres exploiting free-energy gradients from available substrates \citep{Nicholson2022}. Organisms use the available \ch{H2} and \ch{CO} to produce energy (catabolism) via pathways such as:

\begin{equation}
\label{eq:ch6_H2cat}
    \ch{4 H2 + CO2 -> 2 H2O + CH4}\;(+\text{\;energy}),
\end{equation}

\begin{equation}
\label{eq:ch6_COcat}
    \ch{4 CO + 2 H2O -> 2 CO2 + CH3COOH} \;(+\text{\;energy}),
\end{equation}

\noindent where this energy can be used for other metabolic processes such as building biomass (anabolism), represented here as \ch{CH2O}:

\begin{equation}
\label{eq:ch6_H2ana}
    \ch{2 H2 + CO2}\; (+\text{\;energy}) \ch{-> CH2O + H2O},
\end{equation}

\begin{equation}
\label{eq:ch6_COana}
    \ch{2 CO + H2O} \;(+\text{\;energy}) \ch{-> CH2O + CO2}.
\end{equation}

\noindent The biomass will either descend to the ocean floor once the organism has died and is buried, which contributes a burial of reducing matter and oxidises the Atmosphere-Ocean system. Alternatively, secondary consumers would evolve that consume and recycle this biomass:

\begin{equation}
\label{eq:ch6_2ndmet}
    \ch{2 CH2O -> CH3COOH -> CH4 + CO2}.
\end{equation}

\noindent Some fraction of biomass however is inevitably buried, although this percentage is low on Earth with estimates of 0.2\% in modern oceans \citep{Berner1982} and could have been 2\% during the past based on studies on anoxic waters \citep{Arthur1994}. 

On Earth, anoxygenic photosynthesis evolved relatively early, which provided energy directly to create biomass rather than from catabolism reactions. If this were to occur on other planets for \ch{H2} and \ch{CO} consuming organisms, we capture this in the model in the simplified form of

\begin{equation}
    \ch{2 H2 + CO2}\; +\;h\nu \ch{-> CH2O + H2O},
    \label{eq:H2phot}
\end{equation}

\begin{equation}
    \ch{2 CO + H2O} \;+\;h\nu \ch{-> CH2O + CO2}.
\end{equation}

For primary producers, the growth rate is the proportion of total carbon consumed that is fixed into organic matter (\ch{CH2O}). This means that the more carbon that can be converted into organic carbon, the faster an organism will grow. Thus, organisms that obtain energy for carbon fixation from the anabolic metabolism of carbon (Equation~\ref{eq:ch6_H2ana}/\ref{eq:ch6_COana}) have a low growth rate. Faster growth rates can be achieved when the energy for carbon fixation comes from other sources, such as through photosynthesis (Equation~\ref{eq:H2phot}), where the growth rate is close to unity. The implication of this for biosignatures is that organisms with higher growth rates produce less biosignature gases (in this case \ch{CH4}), meaning efficient recycling and thus a low organic carbon burial rate would be required to produce the largest biosignature possible for this ecosystem.

\noindent As abiotic sources of \ch{O2} have been predicted from \ch{CO2} photolysis, its build-up would also provide an energy source for oxic metabolisms, for example by reaction with \ch{CH4} \citep[known on Earth as methanotrophy][]{Kasting2001}:

\begin{equation}
    \ch{2 O2 + CH4 -> 2 H2O + CO2} \;(+\text{\;energy}).
\end{equation}

\noindent Other pathways would be possible depending on what reducing species are most readily available, such as \ch{CO}:

\begin{equation}
    \ch{O2 + 2 CO -> 2 CO2} \;(+\text{\;energy}).
\end{equation}

In this work, we assume that the primary productivity of the biosphere is limited by the supply of electron donors. Prior to the evolution of oxygenic photosynthesis, it is thought that primary productivity was limited by electron donor supply \citep{Ward2019}, rather than phosphorus, nitrogen or the availability of other micronutrients. Oxygenic photosynthesis uses \ch{H2O} and \ch{CO2} to fix carbon and with both of these species in high abundance, phosphorus and nitrogen therefore limit productivity. Therefore, here (in the absence of oxygenic photosynthesis) we assume that the less abundant electron donors, \ch{H2} and \ch{CO}, limit productivity, following previous studies \citep[e.g.][]{Kharecha2005,Ozaki2018,Sauterey2020}. Other limiting factors could have been important on the early Earth and could be important for exoplanets, with ocean transport playing an key role \citep{Olson2020,Salazar2020}. Therefore, this work presents upper limits for biosphere productivity for an assumed flux of abiotic \ch{H2} and \ch{CO}.

\section{Methods}
\label{sec:methods}
In this work, we use the Platform for Atmosphere, Land, Earth and Ocean (PALEO) modelling framework \citep{DainesLenton2016} to study potential abiotic conditions on planets as well as life resembling that of life on Earth prior to the evolution of oxygenic photosynthesis. PALEO relies on the use of a 1D photochemical atmosphere model coupled to a single box ocean that hosts a biosphere. The model conserves redox balance across the atmosphere ocean system. We do these experiments for Earth around 3.8\,Ga, when these ecosystems were thought to be present and for TRAPPIST-1e. TRAPPIST-1e is one of several planets in orbit around TRAPPIST-1 \citep{Gillon2017}, an ultra-cool M-dwarf, with the planets in this system prime targets for atmospheric characterisation. The atmospheric composition and climate solutions for the TRAPPIST-1e simulations are then used to generate synthetic spectra using the Planetary Spectrum Generator (PSG) \citep{Villanueva2018} to predict the potential detectability of biospheres prior to the evolution of oxygenic photosynthesis on TRAPPIST-1e. The components of this model are now described in more detail.

\begin{figure*}
    \centering
    \includegraphics[width=\linewidth]{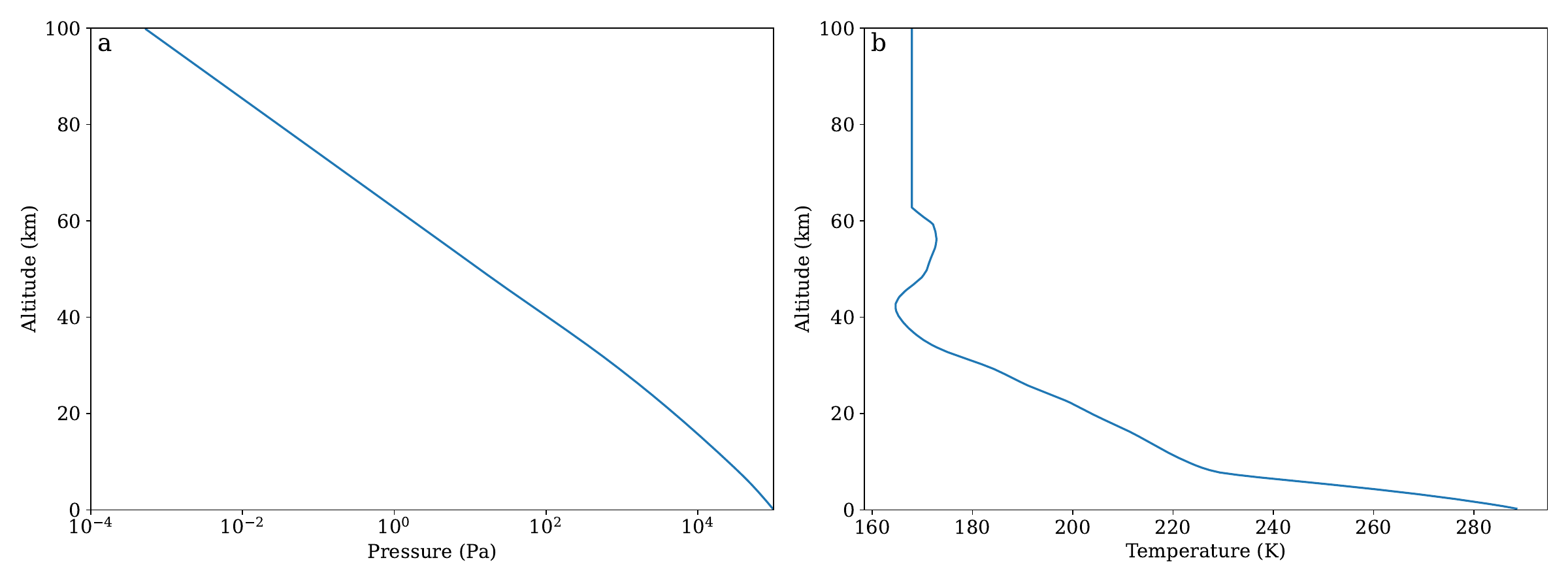}
    \caption{Pressure (a) and Temperature (b) vertical profiles used that were generated for 3.8\,Ga Earth.
    \label{Fig:ptgrid}
    }
\end{figure*}

\subsection{Climate solution}
The 1D atmosphere includes a radiative-convective model to calculate the climate solution that provides a vertical temperature and humidity profile. This climate state is then used by the photochemical model to calculate chemical  rates. For the purpose of this work, these components are not coupled and a single climate solution is used for each planetary configuration, even though atmospheric composition varies.

Radiative transfer is calculated using the Suite Of Community RAdiative Transfer codes based on \citet{EdwardsSlingo1996} (SOCRATES) \citep{Manners2022}, with a convective adjustment based on \citet{ManabeStrickler1964}, with a pseudoadiabatic lapse rate. Tropospheric water vapour decreases with temperature \citep{ManabeWetherald1967} with a prescribed surface relative humidity of 70\,\%. Above the tropopause, the water vapour mixing ratio is fixed at the tropopause value. We obtain a climate solution for the Earth only, using a solar constant of 75\% of the modern value (1036\,W/m$^2$) for the 3.8\,Ga Earth, calculated from \citet{Gough1981}. SOCRATES employs the correlated-k method for radiative transfer. Thermal radiation is treated via 17 bands (between 3.3\,$\mu$m-10\,mm), while stellar radiation is treated by 43 bands (0.20-20\,$\mu$m). These are suitable for atmospheres dominated by a mixture of \ch{N2} and \ch{CO2} (from 1\% to 20\%), with up to 3.5\% \ch{CH4} \citep[see tests in][]{Eager-Nash2023}, supporting surface pressures up to 10$^6$\,Pa. These include \ch{CO2} sub-Lorentzian line wings and \ch{CO2} self-broadening. Collision induced absorption is included for: \ch{N2}--\ch{CH4}, \ch{N2}--\ch{N2} and \ch{CO2}--\ch{CO2} from HITRAN \citep{Karman2019}, and \ch{CH4}--\ch{CO2} from \citet{Turbet2020}. Line data are from HITRAN 2012 \citep{Rothman2013}. The solar spectrum for the climate solution is taken for a 2.9\,Ga Sun spectrum from \citet{Claire2012}, which has little difference in climate compared to a 3.8\,Ga given the same solar constant is used. 

We generate one climate state for all photochemical simulations, which is based on a 3.8\,Ga Earth. We impose a climate with a composition of \ch{N2}, \ch{CO2} and \ch{CH4} for a 1 bar atmosphere. \ch{CO2} is fixed at 10\% to provide similar surface temperatures (Figure~\ref{Fig:ptgrid}). \ch{CH4} is fixed at 0.18\%, while \ch{N2} makes up the remaining atmosphere in each case. The assumption of using a mean climate state is justifiable as long-term modelling including a carbon cycle finds temperatures return to the abiotic steady state \citep[e.g.][]{Sauterey2020}. We do not use climate solutions for TRAPPIST-1e due to the limitations of our model with regards to representing a tidally locked planet. Using the planetary and orbital parameters of TRAPPIST-1e, we obtain a climate solution with a tropopause height that is too low (4.5km), caused by the low solar constant and a realistic climate solution would only be attainable when using a multi-column model to resolve the day and night sides.

\subsection{Atmosphere-Ocean}
The coupled atmosphere-ocean model with a given climate solution solves the continuity equation:

\begin{equation}
    \frac{\partial n_i}{\partial t} = P_i - n_i L_i -\frac{\partial \Phi_i}{\partial z},
    \label{equ:chemcontinuity}
\end{equation}

\noindent where $n_i$ is the number density of species $i$, $P_i$ is the production rate of the species and $L_i$ is the loss rate. $\Phi_i$ is flux of species $i$ from vertical transport. This is solved for the atmosphere-ocean as a whole, with flux across the atmosphere--ocean boundary calculated using a stagnant boundary layer model \citep{LissSlater1974}. PALEO is a flexible modelling tool using a single code, which uses a single solver to model various components of the Earth system together, such as the atmosphere and ocean in this work. Simulations were run for 3 to 800 million Earth years to allow the system to reach steady-state --- when the reductant input is equal to reductant output through the sum of hydrogen escape and organic carbon burial.  We now describe components of this model in more detail.

\subsubsection{Photochemistry}

We use two reaction networks, the full network containing reactions with carbon, hydrogen, oxygen, nitrogen and sulphur species and a reduced network containing just carbon, hydrogen and oxygen species. The species in each network are shown in Table~\ref{tab:boundary_conditions}, with the left side of the table showing the species in the reduced network only. The reduced network is used as it converges more readily, while the larger network is used to validate the results of the reduced network. Some species are assumed to be short-lived, meaning that their chemical lifetime is short enough to assume that the number density can be determined directly from the chemical production and loss rates only. As the inclusion of short lived species can violate mass balance \citep{Hu2012,Harman2015}, we check the simulations for the conservation of atmospheric redox balance. The full network is composed of 399 reactions, compared to 207 in the reduced network, with the network derived from \citet{Gregory2021}, which is shown in Appendix~\ref{ref:network}. Reaction rates are calculated as described in Appendix~\ref{ref:network}. Solar flux is split into 750 bins from 117.65 to 1000\,nm, similar to the ATMOS model \citep{Lincowski2018,Teal2022}. This included the adoption of a two-stream approach to track stellar radiation for photolysis \citep{Teal2022}. Cross-sections and quantum yields come from the ATMOS open access repository\footnote{\url{https://github.com/VirtualPlanetaryLaboratory/atmos}} \citep{Lincowski2018,Teal2022} and include the updated cross-sections for \ch{H2O} \citep{Ranjan2020}. The atmosphere extends to 100\,km with 200 equally-spaced levels. The surface boundary conditions are shown in Table~\ref{tab:boundary_conditions}. As a result of uncertainties in the reductant input, we investigate a range of potential values of \ch{H2} volcanic outgassing, $\Phi_{\mathrm{outgas}}(\ch{H2})$, from 0.1 to 100 Tmol/yr. Fluxes are given in units of Tmol/yr, which are global fluxes equivalent to an Earth sized planet. For the full network, we include a surface boundary flux of \ch{NO} and \ch{CO}, which represents the role of lightning \citep{Harman2018}. In the reduced network, we exclude both of these fluxes as we do not include nitrogen chemistry in this network and the \ch{CO} flux would introduce an additional reductant input, discussed further in Section~\ref{sec:global_redox_balance}.

\begin{table*}
\begin{center}
 \caption{Species list and abiotic surface boundary conditions used in the model, which are either flux based or use a fixed mixing ratio. Fluxes are given in Tmol/yr and photochemical units (pu - molecules/cm$^2$/s), with a conversion rate of 1\,pu equal to $2.7\times10^{-10}$ used here. Species that are short lived are indicated. Dry deposition velocities values are also stated. The CHO network includes species on the left side of the table only.\\ $^{\dag}$CO flux of this value is used for the full network only as part of lightning flux (equal to NO flux). For a sensitivity study in Section~\ref{sec:COtest} a range of abiotic CO fluxes were used from 0.1 to 100\,Tmol/yr, the same range as used for \ch{H2}.}
 \label{tab:boundary_conditions}
 \begin{tabular}{lccccc|lccccc}
  \hline\textbf{Species} & \textbf{Short} & \textbf{Flux} & \textbf{Flux} & \textbf{Mixing}  &   \textbf{Dry} & \textbf{Species} & \textbf{Short} & \textbf{Flux} & \textbf{Flux} & \textbf{Mixing}  &   \textbf{Dry}  \\ 
  \textbf{} & \textbf{lived} & \textbf{(Tmol/yr)} & \textbf{(pu)} & \textbf{ratio}  &   \textbf{deposition} & \textbf{} & \textbf{lived} & \textbf{(Tmol/yr)} & \textbf{(pu)} & \textbf{ratio}  &   \textbf{deposition}  \\ 
  & \textbf{} & \textbf{} & \textbf{} & \textbf{(mol/mol)} & \textbf{velocity} & & \textbf{} &\textbf{} & \textbf{} & \textbf{(mol/mol)} &   \textbf{velocity}  \\
  & \textbf{} & \textbf{} & \textbf{} & \textbf{} & \textbf{(cm/s)} & & \textbf{} &\textbf{} & \textbf{} & \textbf{} &   \textbf{(cm/s)}  \\
  \hline 
  \input{species_list}
 \end{tabular}
 \end{center}
\end{table*}

Tropospheric water vapour is fixed and decreases with temperature \citep{ManabeWetherald1967}, with a prescribed surface relative humidity of 70\,\%. Above the tropopause water vapour may evolve from chemical sources and sinks, but when water vapour exceeds a critical relative humidity (100\,\%), it will condense and rainout.

The species can be deposited at the surface of the atmosphere as a result of either species dissolving into water droplets and subsequently raining out of the atmosphere, or through the particles directly settling from the atmosphere to the surface due to gravity or turbulence, which is known as dry deposition. Rainout follows the prescription described in \citep{GiorgiChameides1985}, while dry deposition is modelled at the lowest atmosphere level, with deposition flux, $\Phi^{dep}$ (molecules/cm$^2$/s) for species $i$ is calculated as

\begin{equation}
    \Phi^{dep}_i = n_i v^{dep}_i,
\end{equation}

\noindent where $v^{dep}$ is the deposition velocity. This represents the free fall of species out of the atmosphere. Eddy diffusivity models the upward vertical mixing of the atmosphere and is calculated for species $i$ as:

\begin{equation}
    \Phi^{eddy}_i = -KN\frac{\partial}{\partial z}\frac{n_i}{N},
\end{equation}

\noindent where $K$ is the eddy diffusion coefficient plotted in Figure~\ref{Fig:eddydiff}, $N$ is the total number density of the atmosphere at height $z$.

Escape and molecular diffusion of molecular and atomic hydrogen follow the procedure outlined in \citet{Hu2012} and \citet{Ranjan2020}. We follow the implementation used in \citet{Hu2012}, which combines diffusion and escape as linked processes. We include the escape of atomic and molecular hydrogen for an \ch{N2}-dominated atmosphere, and use the gravitational acceleration of 9.81 and 9.12 m/s$^2$ for Earth and TRAPPIST-1e simulations respectively, with the latter consistent with values used in other studies \citep[e.g.][]{Fauchez2020}.


\begin{figure*}
    \centering
    \includegraphics[width=\linewidth]{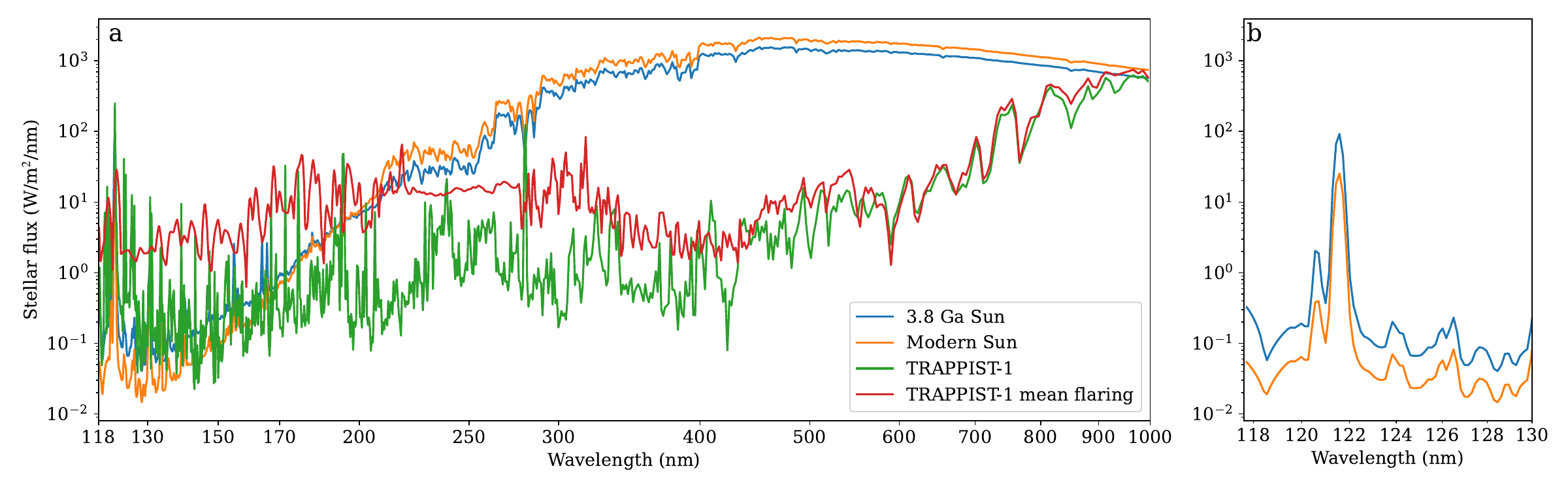}
    \caption{Top-of-atmosphere spectra used for Earth by the modern and 3.8\,Ga Sun and TRAPPIST-1e for a quiescent and mean-flaring TRAPPIST-1 spectrum. (a) shows the spectra for the full range of fluxes for the photochemical model, while (b) shows the flux around Lyman-alpha for the 3.8\,Ga and modern Sun spectra. The spectra from the modern Sun come from the Atmos open repository \citep{Teal2022}, while the 3.8\,Ga Sun spectra comes from \citet{Claire2012}. The TRAPPIST-1 spectra is from \citet{Peacock2019}, while the flaring spectra uses the spectrum from \citet{Wilson2021} and includes a temporally-averaged spectrum which describes a regularly flaring star based on \citet{Ridgway2022}.
    \label{Fig:TOASpectra}
    }
\end{figure*}

The top-of-atmosphere spectra received by the planet used in this work are shown in Figure~\ref{Fig:TOASpectra}. The solar spectrum for 3.8\,Ga is generated from \citet{Claire2012}, while the quiescent spectrum for TRAPPIST-1 uses the spectrum from \citet{Peacock2019}. The mean-flaring spectrum is generated using the same approach described in \citet{Ridgway2022} using the TRAPPIST-1 spectrum from the Mega--MUSCLES survey \citep{Wilson2021}.

\subsubsection{Ocean}

A single box ocean is connected to the atmosphere, with reservoirs of species connected between the atmosphere and ocean via diffusion. A single box ocean is used as we assume a biosphere that is globally spread, which follows approaches used in \citet{Kharecha2005,Ozaki2018,Nicholson2022}. Using a similar procedure to \citet{Kharecha2005} the diffusion is determined by a piston velocity and solubility coefficient, with values given in Table~\ref{tab:diffusion}. The flux across the ocean-atmosphere boundary for species $X$, $\Phi(X)$ is then:
\begin{equation}
    \Phi(X)=v_p(X)\times(\alpha(X)\times pX-[X]_{aq})\times C,
\end{equation}
where $v_p(X)$ and $\alpha(X)$ are the piston velocity and the Henry's solubility coefficient of $X$, respectively, $pX$ is the surface partial pressure of $X$ and $[X]_{aq}$ is the ocean concentration of $X$. $C$ is a constant for unit conversion to obtain a flux in mol/yr.

\begin{table}
\begin{center}
 \caption{Solubility and piston velocity values used within the model.}
 \label{tab:diffusion}
 \begin{tabular}{lcc}
  \hline
  Species & Solubility (mol/m$^3$/Pa) & Piston velocity (m/day) \\
  \hline
  \ch{CO2} & 1.0$\times10^{-2}$ & 4.8$\times10^{-2}$  \\
  \ch{CH4} &  1.4378$\times10^{-5}$ & 4.8 \\
  \ch{H2}  & 8.0106$\times10^{-6}$ & 11.23 \\
  \ch{CO} & 1.027$\times10^{-5}$ & 4.147 \\
  \ch{O2} & 1.2$\times10^{-5}$ & 4.8 \\
  \hline
 \end{tabular}
 \end{center}
\end{table}

In the abiotic simulations and biotic simulations without \ch{CO} consumers, the ocean includes a \ch{CO} sink, following the reaction described in \citep{Harman2015}, taking the form:

\begin{equation}
    \ch{CO + H2O -> H2 + CO2},
\end{equation}

\noindent with the rate of removal of \ch{CO} calculated as $\tau[\ch{CO}]$, where $\tau$ is a timescale of removal, with a value of 10$^{-4}$ used here, and [CO] is the oceanic concentration of CO.

\subsubsection{Biosphere}

We couple a simple ocean with a biosphere component to the 1D atmosphere model, which is shown schematically in Figure~\ref{fig:biosphere_schematic}. The primary production from this biosphere comes from the consumption of \ch{H2} and \ch{CO}, which reach the ocean through the diffusion of these species from the atmosphere. 

The number of mols of \ch{H2}/\ch{CO} required to produce energy for the biomass pathway can be measured experimentally. For methanogens, it has been measured that for every 10\,mols of \ch{CO2} that are metabolised, 1\,mol of \ch{CO2} is converted to biomass through anabolism \citep{Schonheit1980,FardeauBelaich1986,Morii1987}. The proportion of the total metabolism by a species in the form of anabolism is termed the growth rate, $\mu$. For the purposes of this work, it is assumed that anoxic \ch{H2} and \ch{CO} consuming metabolisms have a growth rate of 0.1, while for anoxic photosynthesisers, a growth rate of 1.0 is used as they extract energy directly from the host star to fix carbon. The growth rate has only a minor role in the evolution of the atmosphere, provided the burial fraction is low (Section~\ref{sec:OCB}).

Secondary consumers recycle organic carbon back to \ch{CH4}. However, this process is not 100\,\% efficient and some fraction of the biomass is buried, which may limit the production of \ch{CH4}. The rate of organic carbon burial, $x$, is an unconstrained parameter for exoplanets. However, it has thought to have decreased over Earth history, with a modern value of 0.2\% \citep{Berner1982}. We use a value of 1\% in these simulations unless otherwise stated. This value may have been similar to the Archean, based on measurements of carbon burial in anoxic waters \citep{Arthur1994}. The sensitivity to various recycling rates is analysed in Section~\ref{sec:OCB}, ranging from no recycling to 100\% efficient recycling. The growth rate and organic carbon burial rate are not considered for the oxic metabolisms. However, a sensitivity study of the burial rate and growth rate for the other parameters shows that these only have a minor effect when the burial rate is above 10\%, see Section~\ref{sec:OCB}.

Following similar treatment in \citet{Kharecha2005} and \citet{Ozaki2018}, it is assumed that either \ch{H2} or \ch{CO} are the limiting factors in primary productivity for anoxic metabolisms. Oxic metabolisms are limited by the either availability of \ch{O2} or \ch{CO}/\ch{CH4} depending on the metabolism. The net primary productivity (NPP), in terms of moles of \ch{H2}/\ch{CO} equivalent consumed by the anoxic metabolisms is:

\begin{equation}
    \Phi_{\mathrm{bio}}(\mathrm{NPP}) = \sum_X \tau_{\mathrm{bio}}([X]_{aq}-[X]_{aq}^{lim}),
\end{equation}

\noindent for the sum of limiting species $X$ for each metabolism (\ch{H2} and \ch{CO}) and $[X]_{aq}^{lim}$ is the limiting concentration of species $X$, which the biosphere is unable to draw concentrations below (set here as 4.6$\times$10$^{-7}$\,mol/m$^3$). $\tau_{\mathrm{bio}}$ is the timescale for the biological reaction, which is assumed to be $10^4$\,yr$^{-1}$. Provided this timescale is high enough to reduce ocean concentrations to $[X]_{aq}^{lim}$, which is expected for a global biosphere, this parameter has no effect on the atmospheric configuration. Of this flux of gas consumed some fraction of this flux goes towards producing organic carbon. This proportion is known as the growth rate, $\mu$, where we assume the remaining fraction is used for producing energy to create organic carbon (see Section~\ref{sec:ecosystemprocesses}). Thus, the proportion of organic carbon produced from $\Phi_{\mathrm{bio}}(\mathrm{NPP})$ is

\begin{equation}
    \Phi_{\mathrm{bio}}(\ch{CH2O}) = \frac{1}{2}\mu\times\Phi_{\mathrm{bio}}(\mathrm{NPP}).
\end{equation}

\noindent The factor of half comes from the number of moles of $X$ required to form one mole of organic carbon equivalent (\ch{CH2O}). The burial rate is then:

\begin{equation}
\label{eq:ocb}
    \Phi_{\mathrm{burial}}(\ch{CH2O}) = x\Phi_{\mathrm{bio}}(\ch{CH2O}),
\end{equation}

\noindent where $x$ is the burial rate, the fraction of organic carbon that does not get recycled. The biotic \ch{CH4} flux from these metabolisms is then the difference between these two fluxes:

\begin{equation}
\begin{split}
    \Phi_{\mathrm{bio}}(\ch{CH4})&= \frac{1}{4}\Phi_{\mathrm{bio}}(\mathrm{NPP}) - \frac{1}{2}\Phi_{\mathrm{burial}}(\ch{CH2O})\\ 
    &= \frac{1}{4}(1-x\mu)\Phi_{\mathrm{bio}}(\mathrm{NPP}),
\end{split}
\label{eq:ch4flux}
\end{equation}

\noindent with the fractions used to convert fluxes to \ch{CH4} equivalent fluxes.

In this work, two ecosystems are predominantly used:
\begin{enumerate}
    \item Ecosystem 1: preoxygenic anoxic metabolisms only, including anoxic \ch{H2} and \ch{CO} consumers (Equations 1--4) and secondary consumers (Equation 5)
    \item Ecosystem 2: ecosystem 1 plus oxic metabolisms that metabolise \ch{O2} with \ch{CH4} or \ch{CO} (Equations 8 \& 9)
\end{enumerate}

\subsubsection{Global redox balance}
\label{sec:global_redox_balance}

To conserve global redox balance in our model, we follow similar logic to \citet{Harman2015} to ensure our atmosphere-ocean boundary conserves redox. In our model, the global redox state is conserved by balancing

\begin{equation}
\begin{aligned}
    \Phi_{\mathrm{outgas}}(\mathrm{Red})  = \Phi_{\mathrm{esc}}(\ch{H2}) + 2\Phi_{\mathrm{burial}}(\ch{CH2O}), 
\end{aligned}
\label{equ:global_redox}
\end{equation}

\noindent where $\Phi_{\mathrm{outgas}}(\mathrm{Red})$ is the reductant outgassing through species such as \ch{H2}. $\Phi_{\mathrm{esc}}(\ch{H2})$ is the hydrogen escape flux and $\Phi_{\mathrm{burial}}(\ch{CH2O})$ is the burial of organic carbon, written in terms of its \ch{H2} equivalent redox potential, with reducing fluxes on the left and oxidising fluxes on the right. We do not include in this oxidative weathering or iron and sulphur related burial. Meanwhile redox balance in the ocean is conserved when

\begin{equation}
    \Phi_{\mathrm{dep}}(\mathrm{Red}) = \Phi_{\mathrm{dep}}(\mathrm{Ox}) + 2\Phi_{\mathrm{burial}}(\ch{CH2O}),
\end{equation}

\noindent $\Phi_{\mathrm{dep}}(\mathrm{Ox})$ and $\Phi_{\mathrm{dep}}(\mathrm{Red})$ is the wet and dry deposition of oxidising and reducing species respectively at the atmosphere-ocean boundary. This means that the net deposition at the atmosphere-ocean boundary must equal the organic carbon burial flux. To account for this, the ocean cycling of species that are utilised by the biosphere are separated from the deposition of other species. Species used by the biosphere and abiotic chemistry (Table~\ref{tab:diffusion}) are tracked in the ocean. The net redox flux of these species from biological reactions (through diffusion into and out ocean) is then balanced by the organic carbon burial:

\begin{equation}
    4\Phi_{\mathrm{dep}}(\ch{CH4}) + \Phi_{\mathrm{dep}}(\ch{H2}) + \Phi_{\mathrm{dep}}(\ch{CO}) = 2\Phi_{\mathrm{dep}}(\ch{O2}) + 2\Phi_{\mathrm{burial}}(\ch{CH2O}).
\end{equation}

\noindent The remaining deposition fluxes must then be redox balanced:

\begin{equation}
    \Phi_{\mathrm{dep}}'(\mathrm{Ox}) = \Phi_{\mathrm{dep}}'(\mathrm{Red}),
    \label{equ:surface_redox}
\end{equation}

\noindent where we denote $\Phi_{\mathrm{dep}}'$ as the total deposition fluxes not traced in the ocean. Following the treatment in \citep{Harman2015}, as these fluxes are rarely equal, a balancing flux of \ch{H2} or \ch{O2} is added back into the atmosphere so that this surface exchange is redox neutral. In the case a net reducing deposition surface sink, an equivalent flux \ch{H2} is added back into the atmosphere at the surface, while in the case of a net oxidising surface deposition, an equivalent flux of \ch{O2} is used instead.

To maintain consistency in the reductant input for the full network and the reduced network, surface boundary flux of \ch{H2S} is included in the flux calculation for surface redox balance, such that the net reductant input from \ch{H2S} is zero. This ensures that the net reductant input is determined by the \ch{H2} and \ch{CH4} abiotic fluxes only. \ch{SO2} is considered a redox neutral species, so does not affect the redox balance, while the combined lightning fluxes of CO and NO have an equal and opposite redox state, so also do not contribute to the redox balance.

\subsection{Synthetic spectra}

Synthetic spectra were generated using the Planetary Spectrum Generator (PSG) \citep{Villanueva2018}, adopting the parameters of TRAPPIST-1e \citep{Agol2021}. The following species were included to generate the spectrum: \ch{O2}, \ch{CH4}, \ch{N2}, \ch{C2H6}, \ch{O3}, \ch{CO2}, \ch{CO}, \ch{H}, \ch{H2}, \ch{O}, \ch{C2H2}, \ch{C2H4}. From this, we can investigate the differences in transmission spectrum between biotic and abiotic configurations to assess the detectability of a potential biosphere of the form modelled here. Synthetic observations were then created using PandExo \citep{Batalha2017}. These were generated for the JWST NIRSpec PRISM instrument mode.

\section{Results}
\label{sec:results}

We now present results of abiotic simulations of TRAPPIST-1e, comparing these to simulations of the 3.8\,Ga Earth, showing the importance of reductant input in abiotic \ch{O2} levels. Following this, we present results from the biotic configurations for TRAPPIST-1e, initially benchmarking the coupled atmosphere-ecosystem model against \citet{Kharecha2005} on Earth, and subsequently comparing the differences between early Earth biospheres on Earth and on a hypothetical TRAPPIST-1e. Finally, we show transmission spectra for TRAPPIST-1e for a range of both biotic and abiotic configurations.

\begin{figure*}
    \centering
    \includegraphics[width=\linewidth]{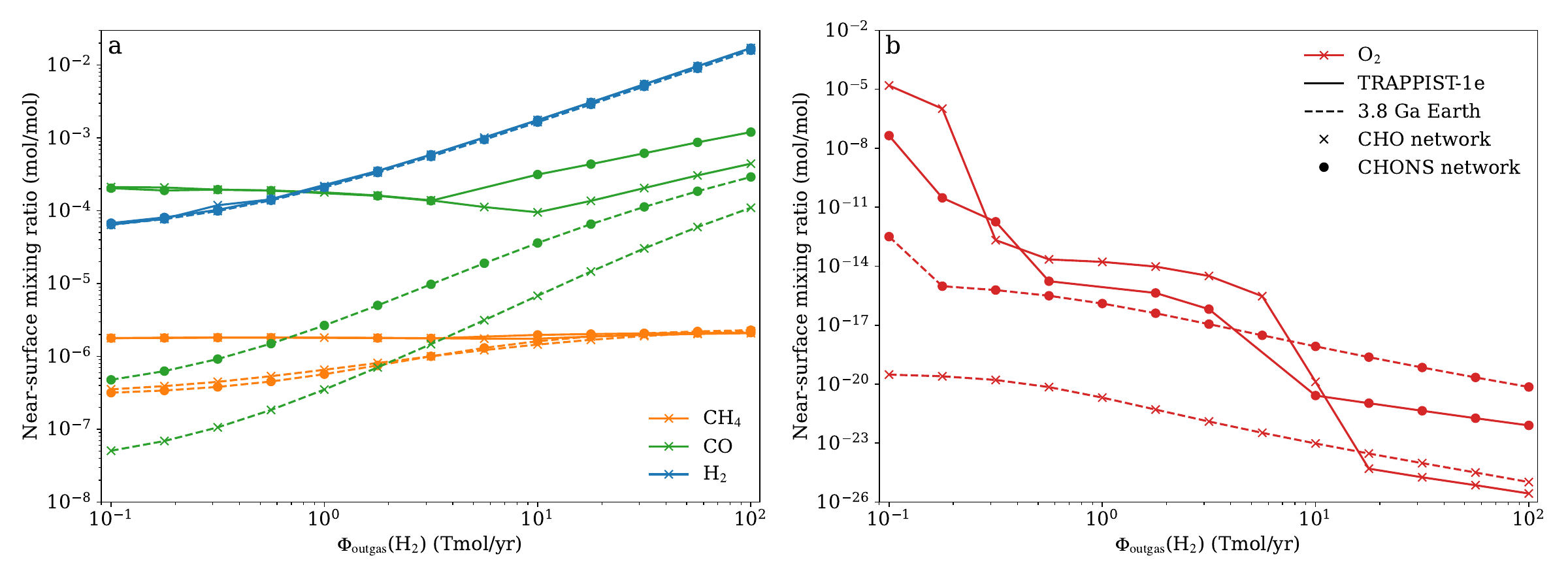}
    \caption{Near-surface (250\,m above the surface) mixing ratios for abiotic configurations of TRAPPIST-1e (solid) and Earth irradiated by a 3.8\,Ga Sun (dashed) as a function of \ch{H2} input. The full grid of simulations is shown for the reduced network (crosses), while a subset of the network is plotted as circles.
    \label{fig:abio}
    }
\end{figure*}

\begin{figure*}
    \centering
    \includegraphics[width=\linewidth]{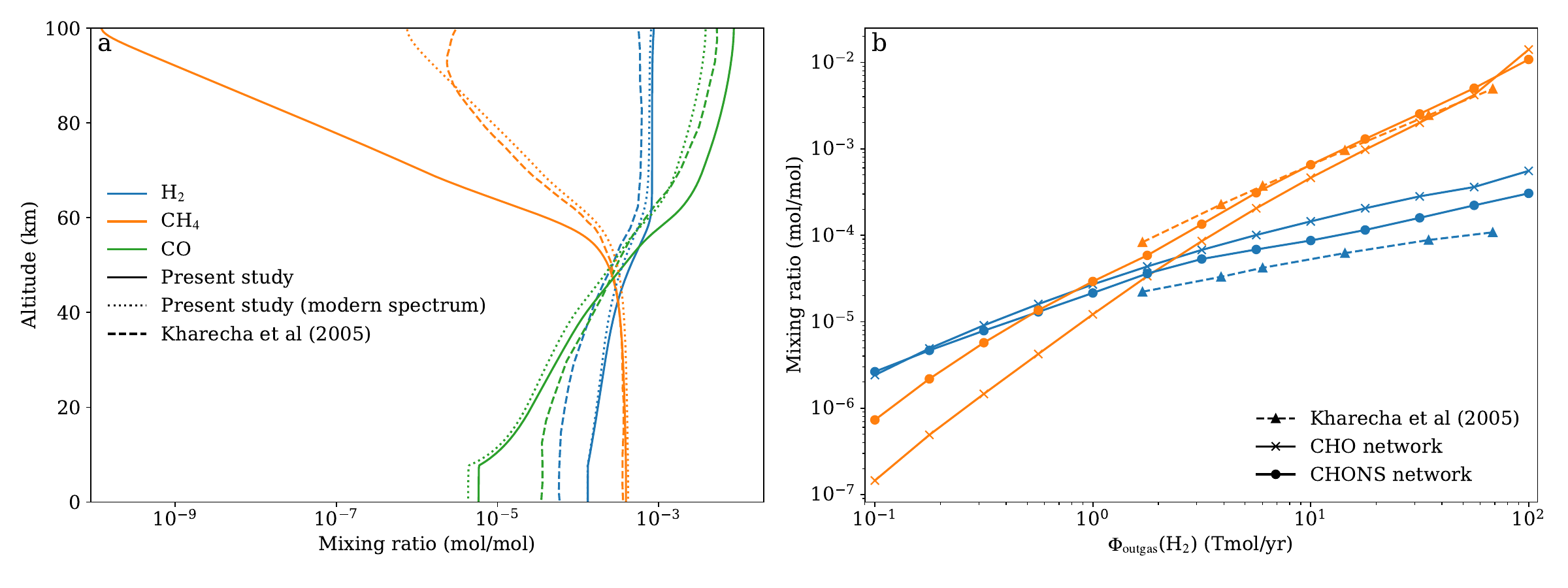}
    \caption{Comparison to \citet{Kharecha2005}. (a) shows mixing ratios vs altitude for case 2 biosphere in \citet{Kharecha2005} (dashed lines), equivalent to their Fig. 7B. This is compared to simulations in this work of the 3.8\,Ga Solar spectrum \citep{Claire2012} (solid line) and the modern Solar spectrum (dotted lines) from ATMOS \citep{Teal2022}. (b) shows the near-surface mixing (250\,m above the surface) ratios of \ch{CH4} and \ch{H2} in this work (solid lines) and data from Table 2 in \citet{Kharecha2005} for their case 3 biosphere (dashed lines). The case 2, (a), and 3, (b), biospheres have growth rates of 0.1 and 1.0 for \ch{H2} consumers respectively, while the CO consumer growth rate is 0 for both cases. The organic carbon burial rate is 2\,\%. The \ch{CO2} near-surface mixing ratio is 2.5\,\%. Our simulations are shown for the reduced network in (a), while (b) shows the full network (circles) and reduced network (crosses).
    \label{fig:kharecha_comp}
    }
\end{figure*}

\subsection{Abiotic \ch{O2} production}

Figure~\ref{fig:abio} shows a comparison of an abiotic configuration of Earth and TRAPPIST-1e, with near-surface mixing ratios of key gases as a function of \ch{H2} input. In the rest of the paper, near surface refers to 250\,m above the surface as this is the lowest atmospheric level. This is shown for the full and reduced network. \ch{H2} is the major hydrogen-bearing species and thus its concentration increases with input, which allows hydrogen escape to balance input. This trend is similar for both Earth and TRAPPIST-1e. \ch{CO} is higher in the TRAPPIST-1e case because of \ch{CO2} photolysis and a lack of tropospheric \ch{OH} from \ch{H2O} photolysis to catalyse the recombination of the \ch{CO2} photolysis products \citep{Harman2015}. Near-surface \ch{O2} is shown in Figure~\ref{fig:abio}b. Reductant input plays an important role in determining abiotic \ch{O2} mixing ratios. At low reductant input, $\Phi_{\mathrm{outgas}}$(\ch{H2})$<10$\,Tmol/yr, \ch{O2} becomes significantly higher for the TRAPPIST-1e case compared to the Earth analogue. This is because reductant input is lower than \ch{O2} production from \ch{CO2} photolysis. When reductant input is higher than \ch{O2} production from \ch{CO2} photolysis, \ch{O2} returns to levels comparable to the Earth case.

\begin{figure*}
    \centering
    \includegraphics[width=\linewidth]{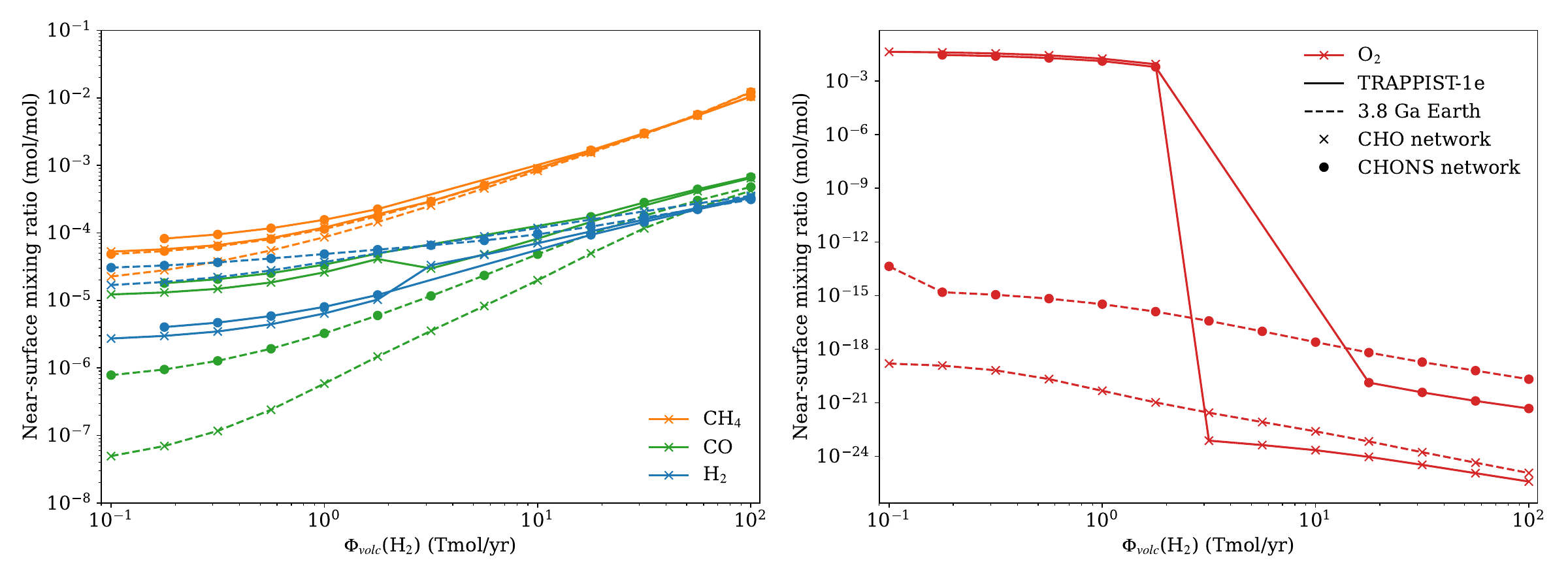}
    \caption{Comparison of a hypothetical \ch{H2} and \ch{CO} consuming ecosystem (ecosystem 1) on TRAPPIST-1e (solid lines) compared to a 3.8\,Ga Earth (dashed lines). (a) and (b) show near-surface (250\,m above the surface) mixing ratios for key gases. The full grid of simulations is shown for the reduced network (crosses), while a subset of the network is plotted as circles.
    \label{fig:sun_trappist_comp}
    }
\end{figure*}

The reduced and full chemical network show similar trends in atmospheric mixing ratio. \ch{H2} and \ch{CH4} remain nearly identical between the two simulations. \ch{CO} is higher for both TRAPPIST-1e (when $\Phi_{\mathrm{outgas}}$(\ch{H2})$>3.16$\,Tmol/yr) and the Earth. This is due to the CO flux from lightning in the full network, which is not included in the reduced network. This is not observed at $\Phi_{\mathrm{outgas}}$(\ch{H2})$\leq3.16$\,Tmol/yr for TRAPPIST-1e as the \ch{CO} source from \ch{CO2} photolysis is much larger. \ch{O2} shows very similar trends between the full and reduced network. For TRAPPIST-1e, the \ch{O2} predictions at $\Phi_{\mathrm{outgas}}$(\ch{H2})$>3.16$\,Tmol/yr are consistent between both networks. For Earth, abiotic \ch{O2} is higher when the full network is used, which produces values closer to those found in \citet{Kasting1990}. The lower values in the reduced network are similar to those quoted in \citet{Ranjan2023}. As these \ch{O2} values are very low, differences are likely to be caused by the numerics of the solver.

\subsection{Benchmarking of atmosphere ecosystem model}
We now consider the effects of a biosphere composed of anoxic metabolisms of \ch{H2} and \ch{CO}. We first compare our model simulations to \citet{Kharecha2005}, who implemented a similar ecosystem model. Figure~\ref{fig:kharecha_comp} shows a comparison to their results for a \ch{H2} and \ch{CO} consuming biosphere with a 2\% organic carbon burial rate. For $\Phi_{\mathrm{outgas}}(\ch{H2})=$6.02\,Tmol/yr (see Figure~\ref{fig:kharecha_comp}a), the mixing ratio profiles of \ch{CH4} are similar up to 70\,km (compare solid and dashed lines), above which we predict much less \ch{CH4}, due to the difference in solar spectrum. We repeat our runs with the modern Earth spectrum (dotted lines), which shows better agreement (compare dotted and dashed lines in Figure~\ref{fig:kharecha_comp}a). The sharper falloff of \ch{CH4} with the 3.8\,Ga spectrum is caused by the high UV flux around Lyman-alpha (Figure~\ref{Fig:TOASpectra}b). The \ch{H2} mixing ratio profiles are similar for both cases, while the \ch{CO} mixing ratio is drawn down to lower values near the surface in our model, but reaches higher abundances at the top-of-atmosphere. Figure~\ref{fig:kharecha_comp}b shows the near-surface mixing ratios of \ch{H2} and \ch{CH4} for their case 3 biosphere, with the \ch{CH4} mixing ratio showing strong agreement with \citet{Kharecha2005}. The difference between the full and reduced reaction network are minimal here, with the only difference at low reductant input, where \ch{CH4} concentrations are marginally higher in the full network. 

\subsection{Pre-photosynthetic biospheres on TRAPPIST-1e}
We now show the effect of a hypothetical \ch{H2} and \ch{CO} consuming biosphere on the atmosphere for a planet orbiting TRAPPIST-1, compared to the young Sun. We show results from ecosystem 1 (anoxic \ch{H2} and \ch{CO} consumers) and subsequently ecosystem 2 (anoxic \ch{H2} and \ch{CO} consumers and oxic \ch{CH4} and \ch{CO} consumers) for simulations with high \ch{O2} mixing ratios. 

\begin{figure*}
    \centering
    \includegraphics[width=\linewidth]{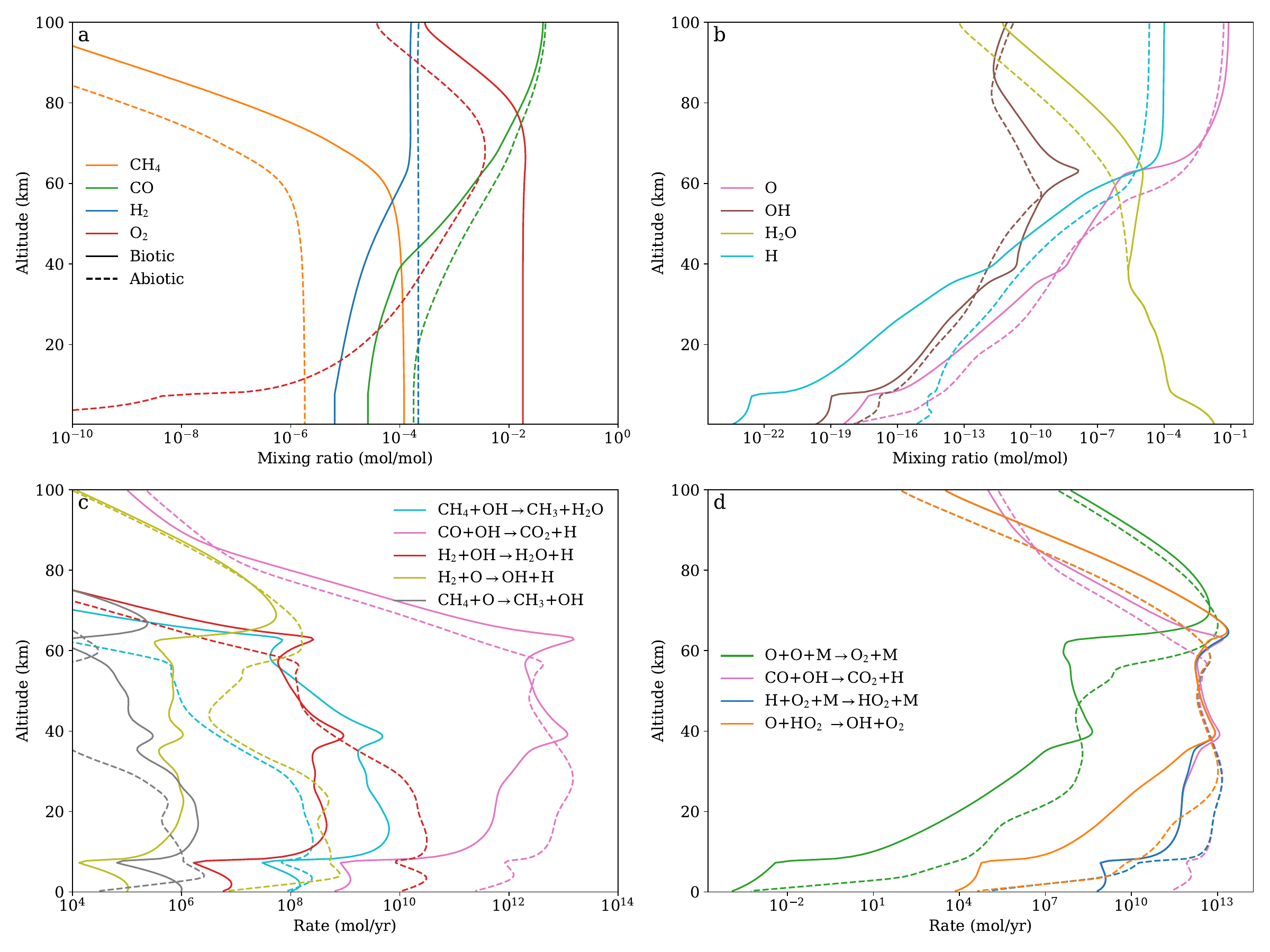}
    \caption{Comparing the ecosystem 1 biotic (solid lines) and abiotic (dashed lines) configuration at $\Phi_{\mathrm{outgas}}(\ch{H2})=$ 1.0\,Tmol/yr, with (a) and (b) showing vertical mixing ratios of selected species and (c) and (d) showing the reaction rates for selected reactions.
    \label{fig:bio_abio_comp}
    }
\end{figure*}

\begin{figure*}
    \centering
    \includegraphics[width=\linewidth]{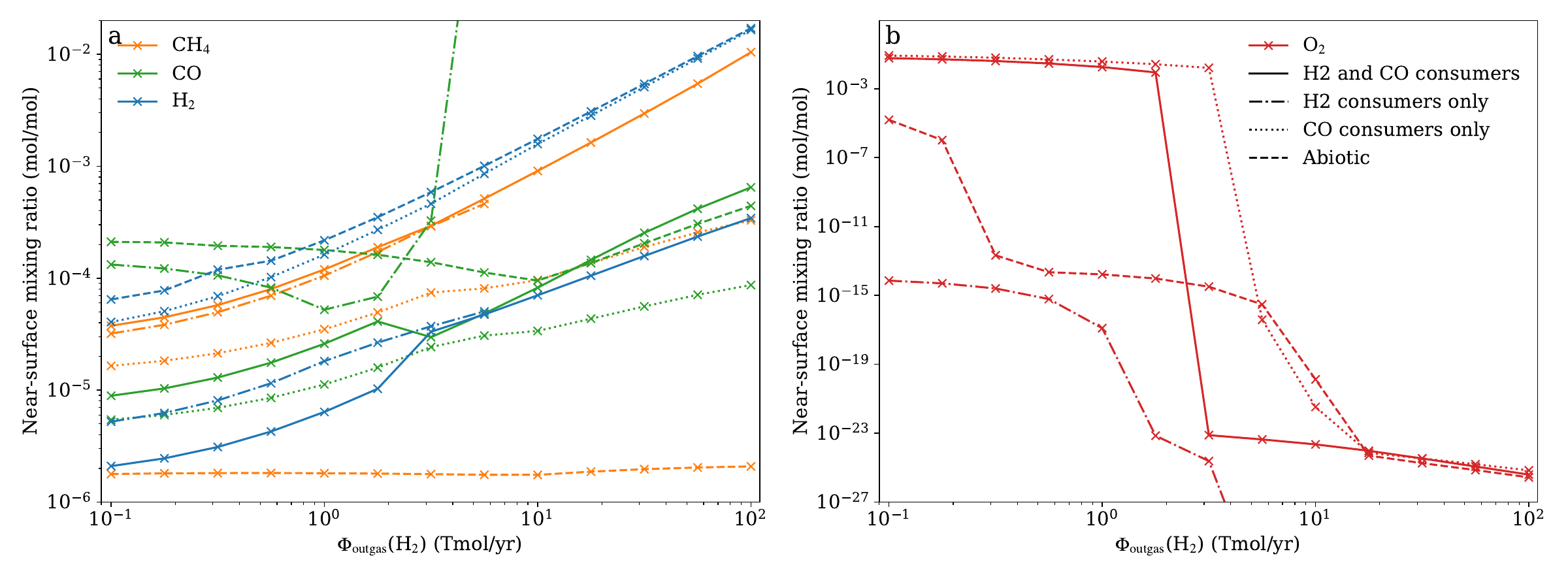}
    \caption{Showing the near-surface (250\,m above the surface) mixing ratios for selected gases for the full ecosystem 1 (\ch{H2} and \ch{CO} consumers, solid lines), \ch{H2} consumers only (dash-dotted), \ch{CO} consumers (dotted) and the abiotic configuration (dashed).
    \label{fig:CompareMetabolisms}
    }
\end{figure*}

\begin{figure*}
    \centering
    \includegraphics[width=\linewidth]{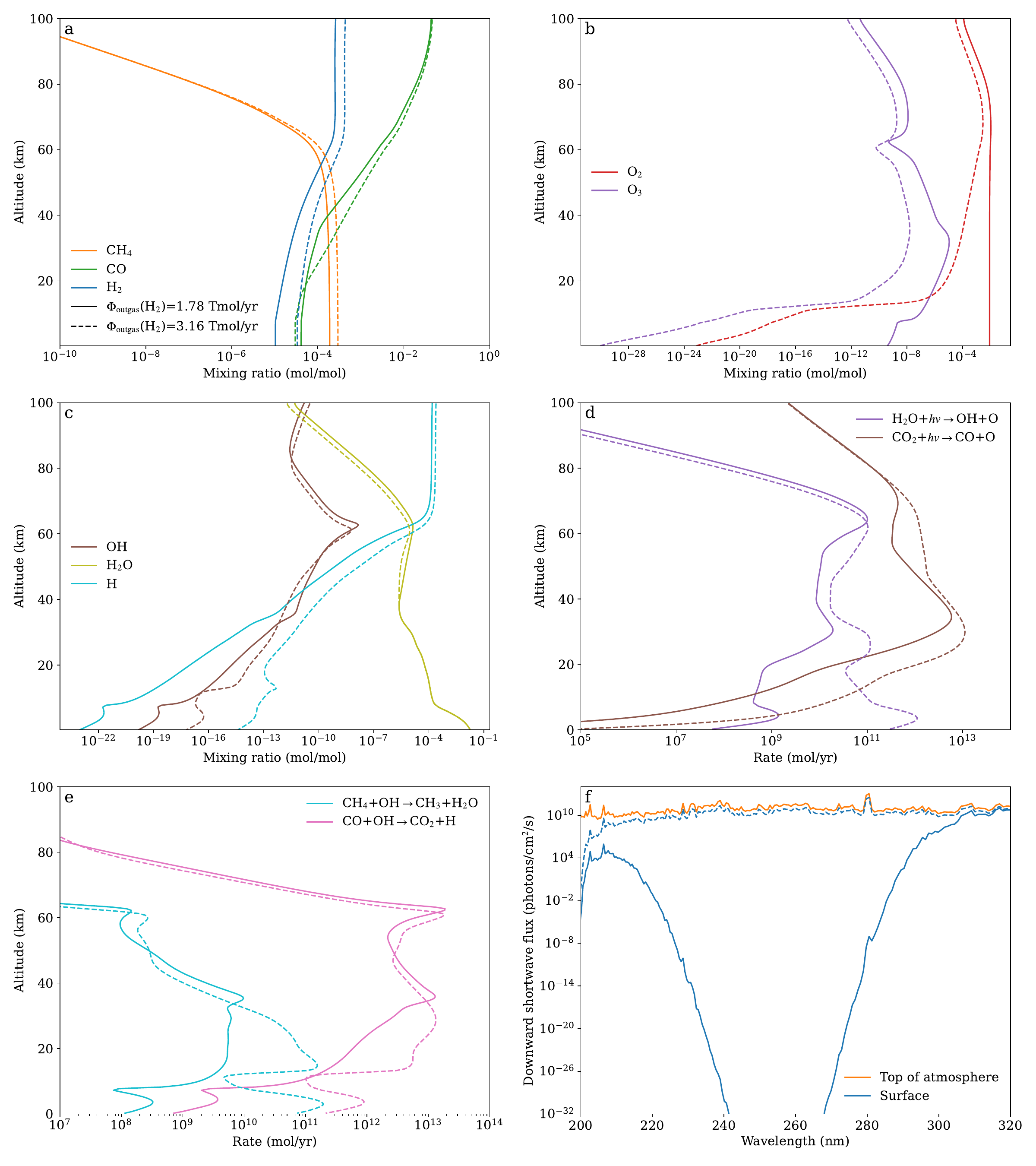}
    \caption{Differences between the low \ch{O2} state at $\Phi_{\mathrm{outgas}}(\ch{H2})=$ 1.78\,Tmol/yr (solid lines) and the high \ch{O2} scenario at $\Phi_{\mathrm{outgas}}(\ch{H2})=$ 3.16\,Tmol/yr (dash-dotted lines). (a), (b) and (c) show gas mixing ratios, while (d) and (e) show reaction rates. (f) shows the downward shortwave flux at the top of atmosphere (TOA) and the surface.
    \label{fig:O2_transition}
    }
\end{figure*}

\begin{figure}
    \centering
    \includegraphics[width=\linewidth]{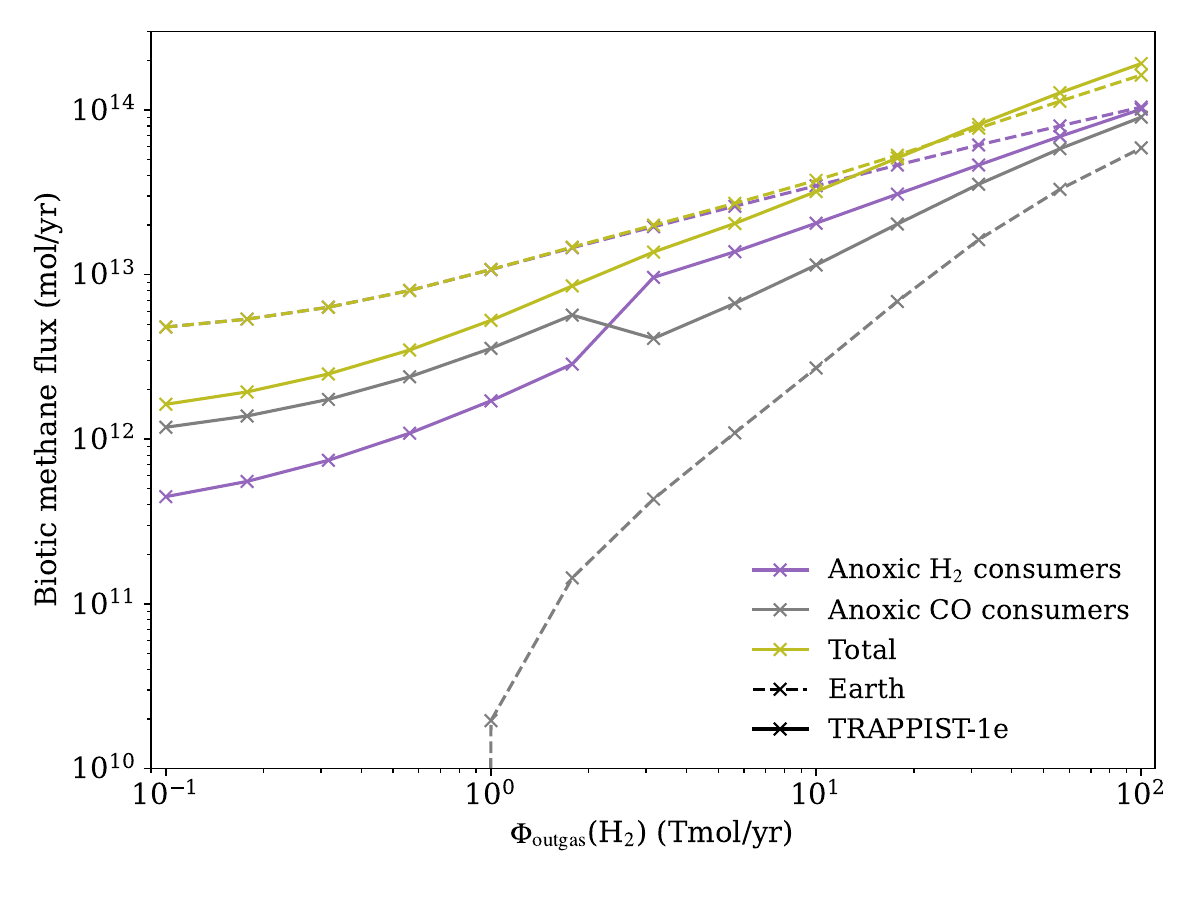}
    \caption{Shows the \ch{CH4} production rate from ecosystem 1 for the 3.8\,Ga Earth (dashed) and TRAPPIST-1e (solid) case. The \ch{CH4} production rate is split between anoxic \ch{H2} (purple) and \ch{CO} (grey) consumers. the total of these is also shown (green) This plot used simulations with the reduced network.
    \label{fig:sun_trappist_prod}
    }
\end{figure}

\subsubsection{Ecosystem 1}
Figure~\ref{fig:sun_trappist_comp} shows the near-surface (250\,m above the surface) mixing ratios for the biotic configuration with ecosystem 1 (anoxic \ch{H2} and \ch{CO} consumers only). The TRAPPIST-1e case has \ch{CH4} at higher concentrations near the surface compared to Earth, which is related to the lower \ch{OH} concentration in the troposphere. The general behaviour of \ch{CH4} is similar for both cases, with \ch{CH4} now the dominant reductant species, and its concentration largely determined by the balance of reductant input with hydrogen escape. The most notable difference however is the relatively high near-surface \ch{O2} (around 1--5\%) at low $\Phi_{\mathrm{outgas}}(\ch{H2})<$ 3.16\,Tmol/yr.

The difference in near surface \ch{O2} mixing ratios between Ecosystem 1 and the abiotic case is caused by the reduction in atmospheric \ch{CO} by \ch{CO}-consuming organisms. The formation of \ch{O2} is predominantly from the combination of \ch{O} in the upper atmosphere (Figure~\ref{fig:bio_abio_comp}d), which originates from \ch{CO2} photolysis. This can oxygenate the atmosphere, when \ch{CO} is consumed by the biosphere. The reduction in atmospheric \ch{CO} due to Ecosystem 1 compared to the abiotic case (Figure~\ref{fig:bio_abio_comp}a) leads to the rate of reaction of \ch{CO} with \ch{OH} dropping by orders of magnitude below 40\,km (Figure~\ref{fig:bio_abio_comp}c). The reaction of \ch{CO + OH} is orders of magnitude higher in the abiotic case compared to reactions of \ch{CH4} and \ch{H2} with \ch{OH} (Figure~\ref{fig:bio_abio_comp}c), and is key to the catalytic loss of \ch{O} \citep{Harman2015}, with associated rates shown in Figure~\ref{fig:bio_abio_comp}d. The oxidising power of \ch{O} in the upper atmosphere is relatively small compared to \ch{OH}, shown by its reaction with \ch{H2} and \ch{CH4} in Figure~\ref{fig:bio_abio_comp}c. The effect of CO-consuming organisms is further demonstrated by the removal of the \ch{CO} metabolism from the Ecosystem 1 simulations.

Figure~\ref{fig:CompareMetabolisms} shows that the removal of \ch{CO}-consuming metabolisms leads to the disappearance of the high \ch{O2} state. The lack of this metabolism leads to \ch{CO} reaching mixing ratios more comparable to the abiotic case, leading to low levels of \ch{O2} (less than 10$^{-14}$\,mol/mol). However, as reductant input increases above 3.16\,Tmol/yr, CO undergoes a runaway, increasing sharply with reductant input. This merits further investigation but requires testing and validation of our model for extreme CO mixing ratios, so we reserve a full investigation for future work. Meanwhile, the presence of \ch{CO}-consuming metabolisms only, without \ch{H2}-consuming metabolisms, retains the original high \ch{O2} state at low reductant input, but with a lower \ch{CH4} and higher \ch{H2} abundance. However, the high \ch{O2} state extends to slightly higher reductant inputs than for when \ch{H2}-consuming metabolisms are present. This is likely due to \ch{CH4} becoming the main \ch{H}-bearing reducing species as opposed to \ch{H2}, meaning the rate of \ch{CO} production increases, which, in turn, decreases the reductant input required to reduce the high \ch{O2} state. As with the CO runaway, this requires further detailed investigation which we will perform in a follow up study.

The high \ch{O2} disappears when reductant input increased compared to the \ch{O2} production from \ch{CO2} photolysis. Figure~\ref{fig:O2_transition} compares the differences between this oxidised state at $\Phi_{\mathrm{outgas}}(\ch{H2})=$ 1.78\,Tmol/yr and the more familiar low \ch{O2} state at $\Phi_{\mathrm{outgas}}(\ch{H2})=$ 3.16\,Tmol/yr. At $\Phi_{\mathrm{outgas}}(\ch{H2})\leq$ 1.78\,Tmol/yr, the high levels of \ch{O2} is the product of \ch{CO2} photolysis followed by the combination atomic oxygen:
\begin{equation}
    \ch{CO2 + } h\nu \ch{ -> CO + O},
\end{equation}
\begin{equation}
    \ch{O + O + M -> O2 + M}.
\end{equation}
When the reductant input is low, \ch{O2} accumulates faster than it can be reduced by the reductant input, leading to an accumulation of \ch{O3} (dotted lines in Figure~\ref{fig:O2_transition}b). This \ch{O3} shields the tropospheric UV flux (at $\Phi_{\mathrm{outgas}}(\ch{H2})=$ 1.78\,Tmol/yr) and reduces the photolysis of \ch{H2O} via:
\begin{equation}
    \ch{H2O + } h\nu \ch{ -> OH + H},
\end{equation}
\noindent which further reduces the loss mechanisms for \ch{O2} as well as \ch{CO} and \ch{CH4}. The high levels of \ch{O2} in the atmosphere lead to low levels of \ch{H2}, causing the low \ch{H2}-consumer productivity at these fluxes with \ch{CO}-consumers contributing the largest \ch{CH4} flux, shown in Figure~\ref{fig:sun_trappist_prod}.

When $\Phi_{\mathrm{outgas}}(\ch{H2})\geq$ 3.16\,Tmol/yr, the reductant input is large enough to significantly reduce the \ch{O2} concentration to levels seen in the Earth case. The subsequent lack of \ch{O3} no longer shields the troposphere from UV radiation, increasing \ch{H2O} photolysis in this region. Increased \ch{OH} leads to lower \ch{CO} and \ch{CH4} abundances. The reduction of \ch{O2} down to levels that are comparable to the Earth case allows for a higher \ch{H2} abundance, which consequently leads to an increase in \ch{H2} consumption, which becomes the dominant source of biological \ch{CH4} production over \ch{CO} consumption (Figure~\ref{fig:sun_trappist_prod}).

The overall biotic \ch{CH4} flux is generally smaller on TRAPPIST-1e compared to the Earth (Figure~\ref{fig:sun_trappist_prod}), which can be attributed to the slower rate of photochemically driven destruction of \ch{CH4} via \ch{OH} \citep{Segura2005}. At low reductant inputs ($\Phi_{\mathrm{outgas}}(\ch{H2})<$10\,Tmol/yr) the \ch{CH4} flux is significantly larger than the reductant outgassing rate, showing the importance of atmospheric recycling of \ch{CH4} back to \ch{H2} and \ch{CO}. As this process is slower for TRAPPIST-1e, the biological \ch{CH4} flux is lower, while the atmospheric \ch{CH4} concentration is up to a factor of two larger. The \ch{CH4} production from CO consumers is more comparable to the H2 consumers as a result of the higher mixing ratios of \ch{CO} on TRAPPIST-1e compared to the Earth (Figure~\ref{fig:sun_trappist_comp}a). When \ch{O2} is high, the \ch{CH4} flux from CO consumers is higher than for \ch{H2} consumers.

The full and the reduced reaction network, both show very similar predictions for the atmospheric composition. This is shown in Figure~\ref{fig:sun_trappist_comp}. Most significantly, the high \ch{O2} state is reproduced with both configurations. As we have now shown that the reduced and full network show very similar results for the abiotic configurations, the reproduction of \citet{Kharecha2005} and for ecosystem 1 for TRAPPIST-1e and the Earth, moving forward we use the reduced network only.

\subsubsection{Ecosystem 2}

\begin{figure*}
    \centering
    \includegraphics[width=\linewidth]{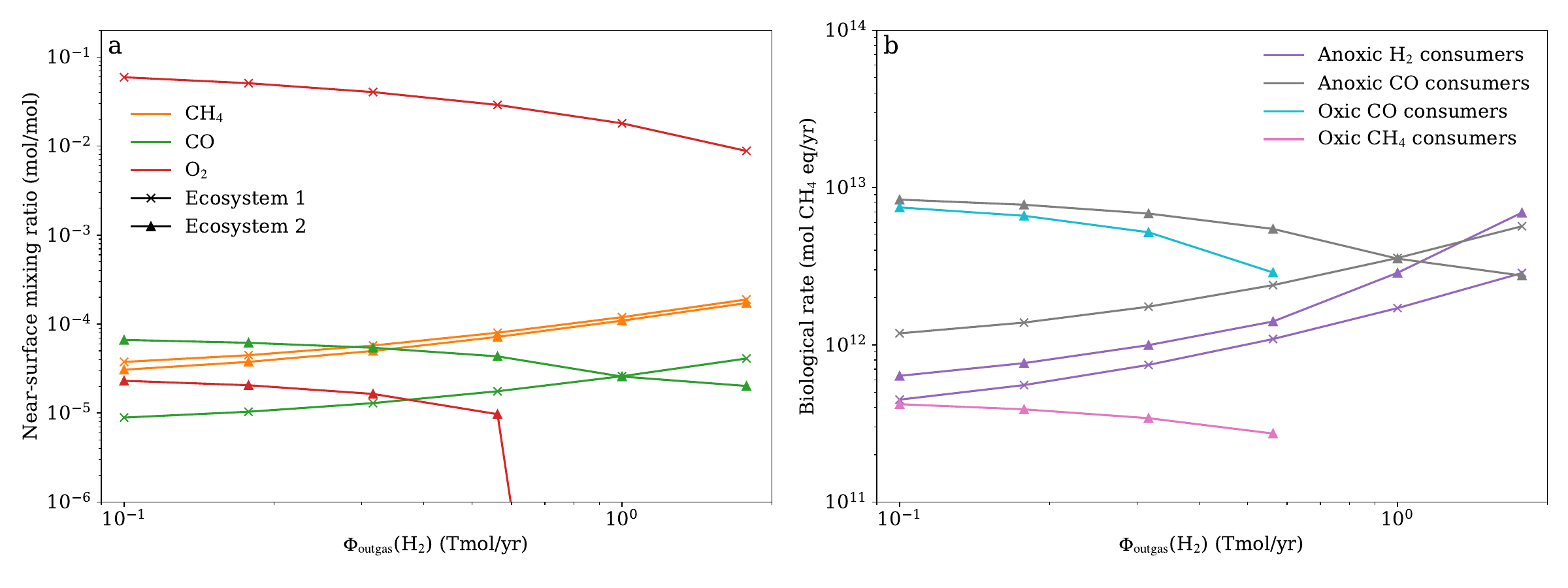}
    \caption{Comparison between ecosystem 1 and 2 at low reductant inputs, which produced high \ch{O2} concentrations. (a) shows the near-surface (250\,m above the surface) mixing ratios for key gases for ecosystem 1 (crosses) and ecosystem 2 (triangles). (b) shows the rate of biological reactions from ecosystem 2 in units of moles of \ch{CH4} equivalent - in terms of \ch{CH4} production for anoxic reactions and in terms of reductant consumption (as either \ch{CO} or \ch{CH4}) in the oxic reactions. Above $\Phi_{\mathrm{outgas}}$(\ch{H2})$=1.78$\,Tmol/yr, the \ch{O2} concentration drops and oxic metabolisms can no longer function, thus becomes identical to ecosystem 1 and is not shown.
    \label{fig:ecosystem2}
    }
\end{figure*}

\begin{figure*}
    \centering
    \includegraphics[width=\linewidth]{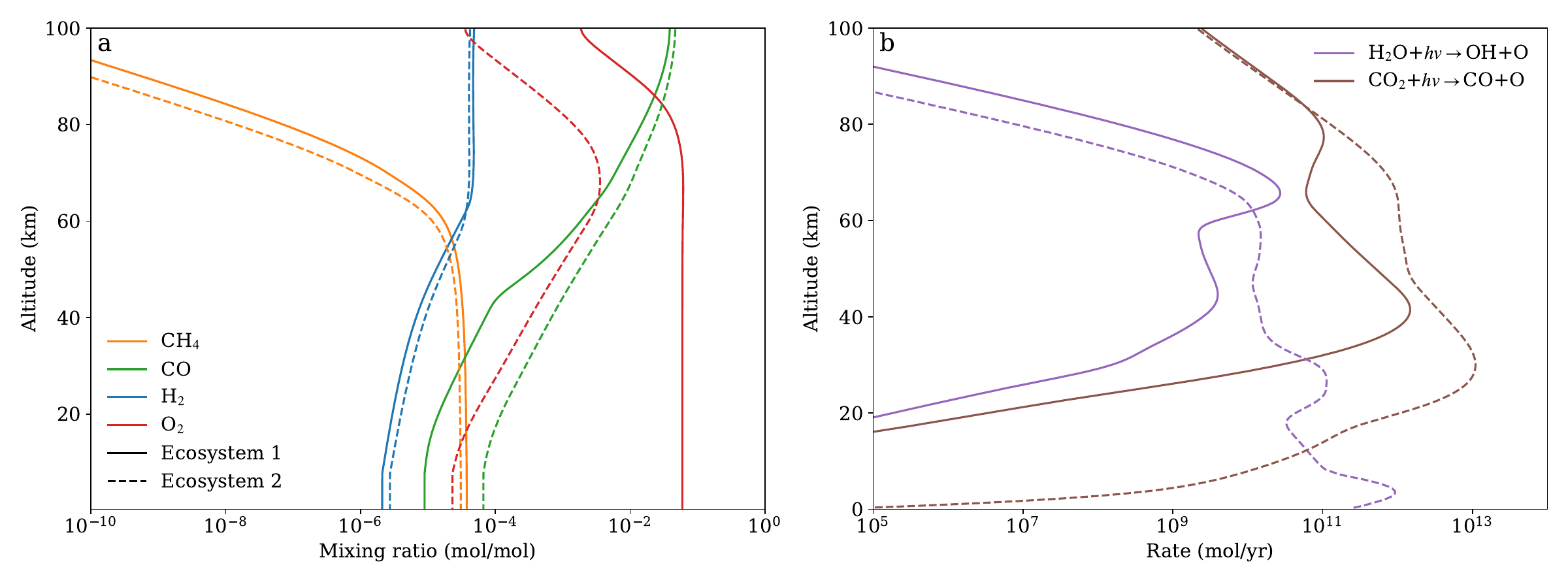}
    \caption{Comparison between the atmospheric composition (a) and photolysis rates (b) produced by ecosystem 1 (solid) and ecosystem 2 (dashed) when $\Phi_{\mathrm{outgas}}$(\ch{H2})$= 0.1$\,Tmol/yr.
    \label{fig:eco1vs2}
    }
\end{figure*}

The presence of high abundances of \ch{O2} provides an energy source for potential ecosystems. The effect of including oxic \ch{CH4} and \ch{CO} consumers alongside the metabolisms in Ecosystem 1 (Ecosystem 2) is shown in Figure~\ref{fig:ecosystem2} for $\Phi_{\mathrm{outgas}}(\ch{H2})\leq$ 1.78\,Tmol/yr. Figure~\ref{fig:ecosystem2}a shows that \ch{O2} concentrations could be drawn down to near-surface mixing ratios of the order $10^{-5}$, whilst also drawing down \ch{CH4} concentrations slightly. However, \ch{CO} is found to rise by up to an order of magnitude (Figure~\ref{fig:ecosystem2}a and Figure~\ref{fig:eco1vs2}a). In this ecosystem, the anoxic and oxic CO consumers dominate the biological productivity, as the oxic \ch{CH4} and  anoxic \ch{H2} consumer productivity is low, shown in Figure~\ref{fig:ecosystem2}b. The productivity of anoxic \ch{CO} consumers is higher than in Ecosystem 1 due to the increased atmospheric \ch{CO} concentration (Figure~\ref{fig:ecosystem2}a).

The rise of atmospheric \ch{CO} is a result of an increase in the photolysis rate of \ch{CO2}, caused by the reduction in \ch{O2} and thus \ch{O3} abundance, shown in Figure~\ref{fig:eco1vs2}b at $\Phi_{\mathrm{outgas}}(\ch{H2})=$ 0.1\,Tmol/yr. The higher \ch{CO} concentration in the atmosphere also contributes to the drop in \ch{O2}, but the presence of \ch{CO} consuming species, which consume more \ch{CO} than \ch{O2} (anoxic CO consumers have a higher rate than oxic CO consumers,shown in Figure~\ref{fig:ecosystem2}), prevents \ch{O2} from falling further. The reduction in \ch{O2} also leads to an increase in the production of tropospheric \ch{OH}, which contributes to the lower \ch{CH4} abundance in Ecosystem 2. At a high enough reductant input, $\Phi_{\mathrm{outgas}}(\ch{H2})>$ 0.56\,Tmol/yr, the presence of oxic metabolisms is sufficient to reduce \ch{O2} to less than $10^{-19}$\,mol/mol. Here, the biological rates of these metabolisms also go to zero, showing the sensitivity of this regime.

\begin{figure*}
    \centering
    \includegraphics[width=\linewidth]{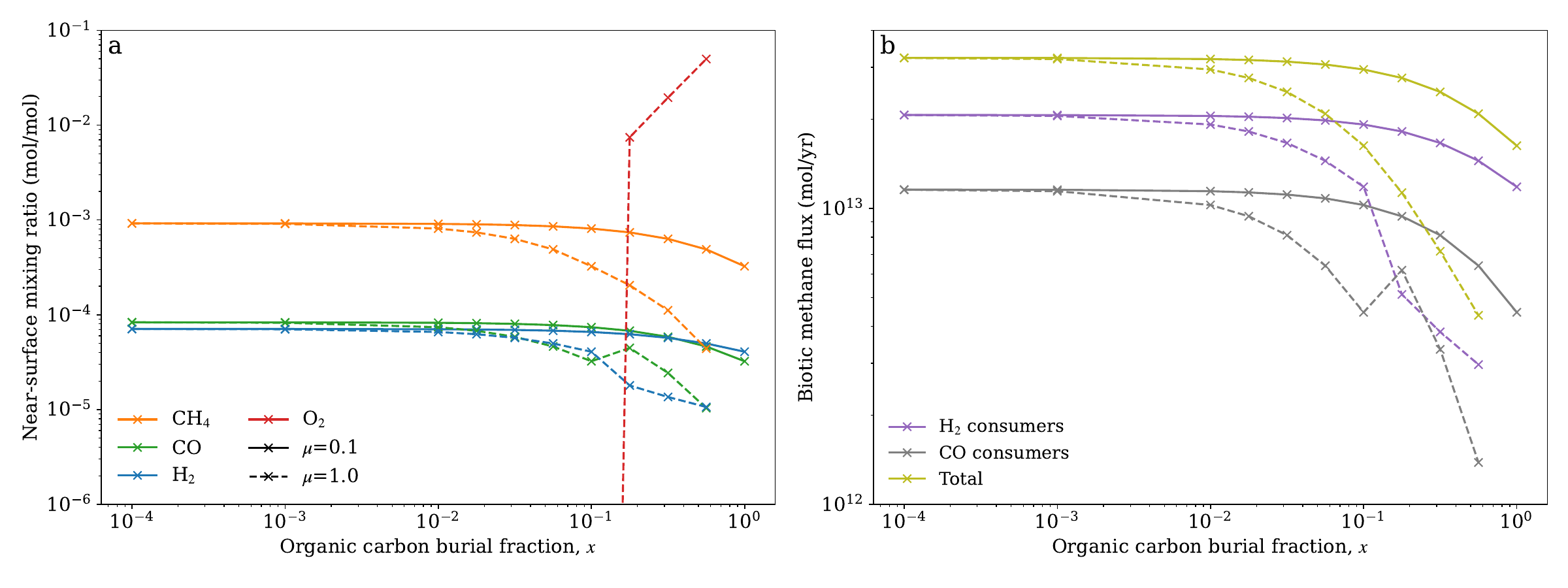}
    \caption{Comparison of the effects of \ch{H2} organic carbon burial rates and growth efficiencies for a \ch{H2} and \ch{CO} consuming biosphere. (a) shows the change in near-surface composition vs burial fraction for \ch{H2}, \ch{CH4}, \ch{CO} and \ch{O2}, while (b) shows the change in biosphere productivity of \ch{H2} and \ch{CO} consuming organisms. $\mu=1.0$ (solid lines) represent a growth rate equivalent to a phototrophic primary production, while the lower growth rate of $\mu=0.1$ (dashed lines) represents a biosphere that produces energy from catabolic reactions instead of using light.
    \label{fig:OCB_sensitivity}
    }
\end{figure*}

\begin{figure*}
    \centering
    \includegraphics[width=\linewidth]{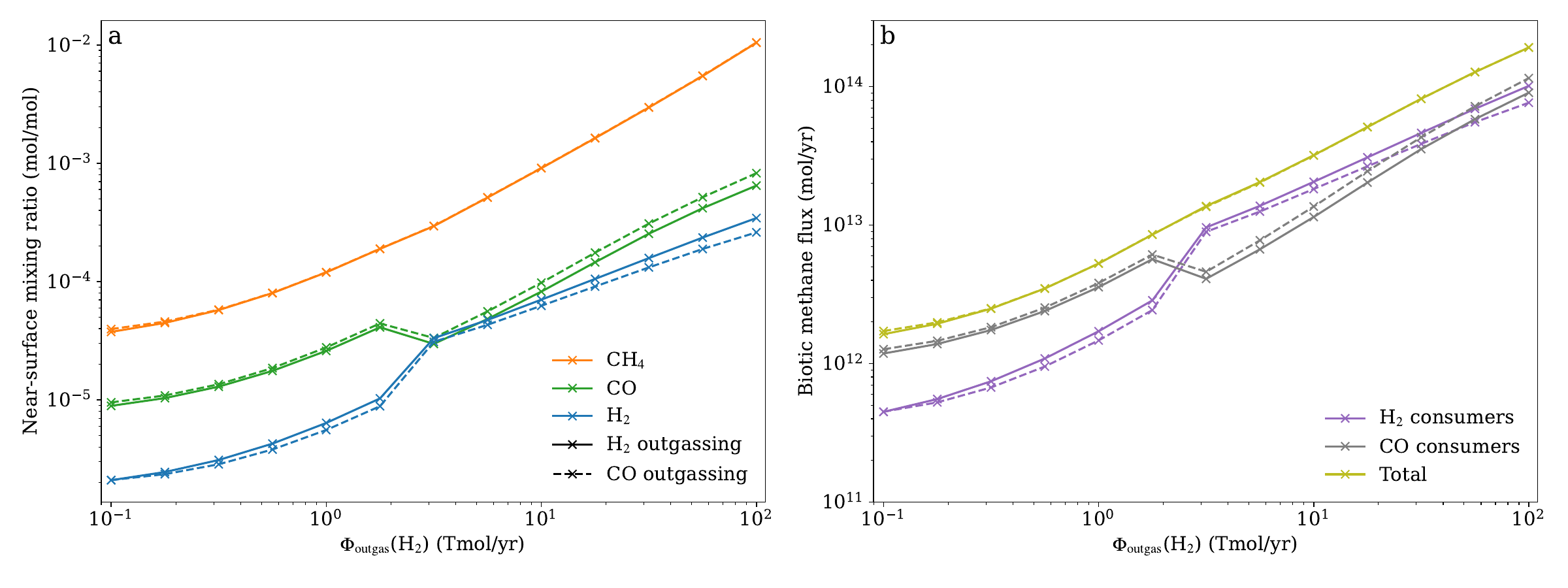}
    \caption{The effects of \ch{H2} outgassing (solid lines) with an equivalent \ch{CO} outgassing. (a) shows the change in near-surface (250\,m above the surface) composition vs the outgassing rate for \ch{H2}, \ch{CH4} and \ch{CO}, while (b) shows the productivity of \ch{H2} and \ch{CO} consuming organisms.
    \label{fig:CO_H2_flux_comp}
    }
\end{figure*}

\subsubsection{Anoxygenic photosynthesis and carbon burial}
\label{sec:OCB}

We now consider the potential impacts of anoxygenic photosynthesis compared to ecosystems that obtain energy through catabolism. That is to say that the growth rate becomes 100\%, as opposed to the 10\% that was assumed previously. To consider the effects of this, we investigate the change in growth rate alongside the sensitivity to the rate of organic carbon burial. We find that anoxygenic photosynthesis using \ch{H2} and \ch{CO} is unlikely to lead to significant changes in the atmospheric composition unless the organic carbon burial rate is low.

Ecosystems with low growth rates, such as prephotosynthetic biospheres are relatively unaffected by the fraction of biomass that is buried. For a growth rate, $\mu$ of 0.1 (10\,\%), we see a change in \ch{CH4} mixing ratio from $\approx$0.001\,mol/mol for a low burial fraction down to $\approx$0.0004\,mol/mol when all organic carbon is buried (Figure~\ref{fig:OCB_sensitivity}a), with the \ch{CH4} mixing ratio only dropping noticeably below 0.001\,mol/mol when the burial fraction is greater than 0.1. This relatively small change is due to the fact that \ch{CH4} is produced via catabolic reactions for 90\% of the \ch{H2} or \ch{CO} that is consumed, while the burial fraction affects what happens to just the remaining 10\,\%. This is further demonstrated in Equation~\ref{eq:ch4flux}, if $x=1$ and $\mu=0.1$, the methane flux will still come from 90\% of the NPP. Adjusting the burial rate can thus only change the methane flux to be 90-100\% of the NPP for this low growth rate case. Increasing the burial rate also has the effect of decreasing the biospheres productivity, shown in Figure~\ref{fig:OCB_sensitivity}b. This is because as the organic carbon burial fraction increases, less \ch{CH4} is returning to the atmosphere, and thus less \ch{CH4} is recycled back to \ch{H2} and \ch{CO}, where it can be consumed by the biosphere again.

In contrast to this, ecosystems with high growth rates, such as via anoxygenic photosynthesis are more sensitive to the organic carbon burial fraction. At low organic carbon burial fractions ($\lessapprox0.01$), the atmospheric composition is  indistinguishable between low and high growth rates, which was also found in Earth like configurations \citep{Kharecha2005}. As the organic carbon burial fraction continues to decrease beyond 0.01, atmospheric \ch{CH4} drops faster than the low growth rate case. This is because when the growth rate is higher, \ch{CH4} is produced entirely by secondary consumers, rather than through catabolism processes, and the burial rate effectively determines how efficient secondary consumers are at recycling biomass. Again, by looking at Equation~\ref{eq:ch4flux}, and this time setting $\mu=1$, now the methane flux can vary from 0 to 100\,\% of the NPP depending on the burial rate, and the system is much more sensitive to this parameter when the growth rate is high. Thus, efficient recycling of organic carbon by secondary organisms is required to produce \ch{CH4} levels that are distinguishable from abiotic configurations (compare Figure~\ref{fig:OCB_sensitivity}a to Figure~\ref{fig:abio}).

For photosynthetic anoxic biospheres, if the organic carbon burial fraction is high, high \ch{O2} scenarios become more likely. For $\Phi_{\mathrm{outgas}}(\ch{H2})=$ 10\,Tmol/yr, a burial fraction of greater than 0.1, can lead to high \ch{O2} states. This is because of the large burial of reductant material and the significantly lower reducing power of the atmosphere from the low \ch{CH4} production, alongside a \ch{H2} consumption from the biosphere.

\begin{figure*}
    \centering
    \includegraphics[width=\linewidth]{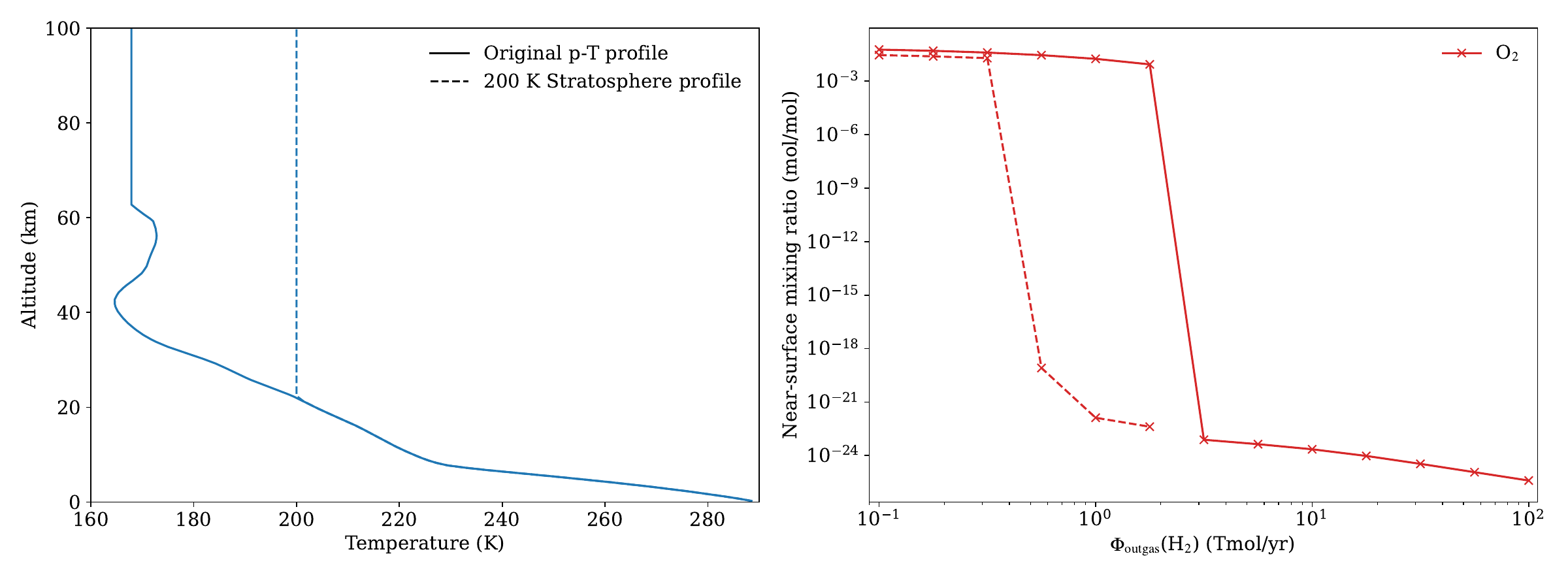}
    \caption{The effect of adjusting stratospheric temperature manually to a minimum of 200\,K (a) on the near surface \ch{O2} mixing ratio (b). Solid lines show the original temperature profile, while the dashed lines show results from the adjusted profile
    \label{fig:strat_adjust}
    }
\end{figure*}

\subsubsection{Carbon monoxide outgassing}
\label{sec:COtest}

As volcanic input can vary depending on the carbon-to-hydrogen ratio in the mantle, we test the sensitivity of the atmosphere to \ch{CO} and \ch{H2} outgassing. Instead of volcanic flux from \ch{H2}, this is replaced with an equivalent \ch{CO} flux, which is found to have a minor effect on the atmospheric composition. Figure~\ref{fig:CO_H2_flux_comp}a shows the near-surface mixing ratios from the key components of the atmosphere. Generally the atmospheric composition is the same for a given \ch{CO} or \ch{H2} flux, particularly for \ch{CH4}, while \ch{H2} and \ch{CO} are similar, apart from at higher volcanic outgassing fluxes where small differences emerge. Figure~\ref{fig:CO_H2_flux_comp}b shows the biotic \ch{CH4} flux associated with these different outgassed gases, which generally show the same trend for both \ch{CO} and \ch{H2} consumers, with \ch{H2} consumer productivity marginally higher when \ch{H2} is outgassed compared to \ch{CO}, and vice versa for \ch{CO} consumer productivity.

\subsubsection{Stratospheric temperature}

The result of high oxygen mixing ratios under low reductant input caused by a CO consuming biosphere is sensitive to the stratospheric temperature. A warmer stratosphere (for example by setting the minimum stratospheric temperature to 200\,K in Figure~\ref{fig:strat_adjust}a) leads to an increase in the stratospheric water vapour content and thus increases the rate of stratospheric water vapour photolysis. This can then increase the rate of recombination of \ch{CO} and \ch{O}, which leads to the reductant limit on \ch{O2} accumulation to occur at a lower reductant input, shown in Figure~\ref{fig:strat_adjust}b.

\subsection{Detectability of pre-photosynthetic biospheres on TRAPPIST-1e}

Finally, we consider the detectability of potential preoxygenic-photosynthesising biospheres for a planet like TRAPPIST-1e. Abiotic and biotic configurations are now presented from PSG. 

\begin{figure*}
    \centering
    \includegraphics[trim={3cm 4cm 3cm 4cm},clip,width=\linewidth]{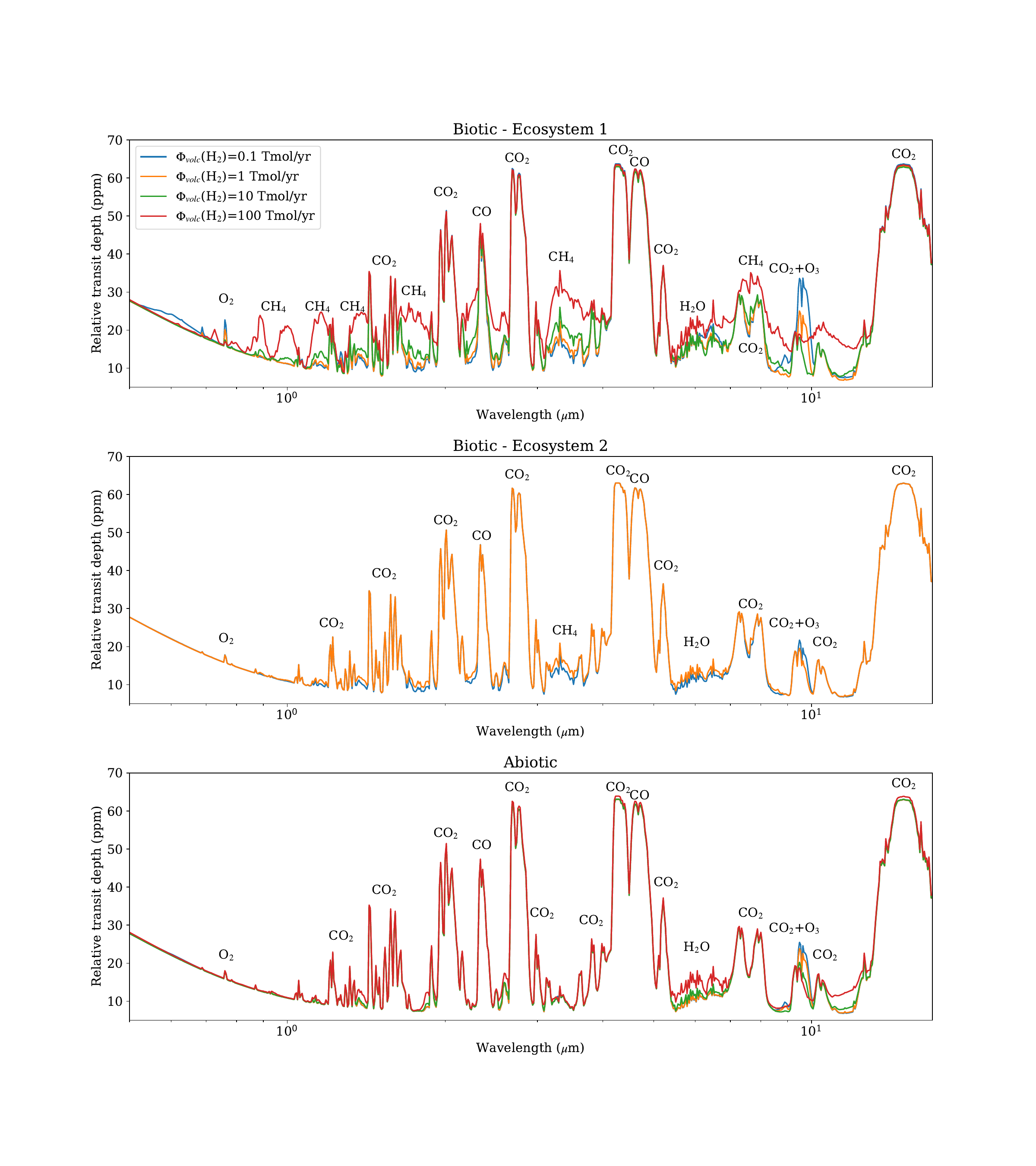}
    \caption{Transmission spectra for biotic and abiotic configurations used here for TRAPPIST-1e with stellar spectra that is quiescent and a mean flaring spectra. Transmission spectra is generated using PSG \citet{Villanueva2018}, with features labelled with the species causing them.
    \label{fig:spectra}
    }
\end{figure*}

Biologically produced \ch{CH4} shows strong signatures at high reductant input, but at low reductant input both the \ch{CH4} and \ch{O2}/\ch{O3} signatures are weak. The transmission spectra for abiotic, and biotic ecosystem 1 and 2 configurations are shown in Figure~\ref{fig:spectra} for a range of $\Phi_{\mathrm{outgas}}$(\ch{H2}). Both \ch{CO} and \ch{CO2} show strong features across all transmission spectra, while \ch{CH4} shows obvious signals only for the biotic configurations, which are most visible at $\Phi_{\mathrm{outgas}}(\ch{H2})=$100\,Tmol/yr. Thus planets may need to have high outgassing rates for these biospheres to be detectable. The strongest \ch{O3} feature is masked by \ch{CO2} at 9.4 microns, although is most visible for ecosystem 1 when the reductant input is lowest. The inclusion of oxic metabolisms leads to this ozone feature being very difficult to distinguish from \ch{CO2}.

\begin{figure*}
    \centering
    \includegraphics[width=\linewidth]{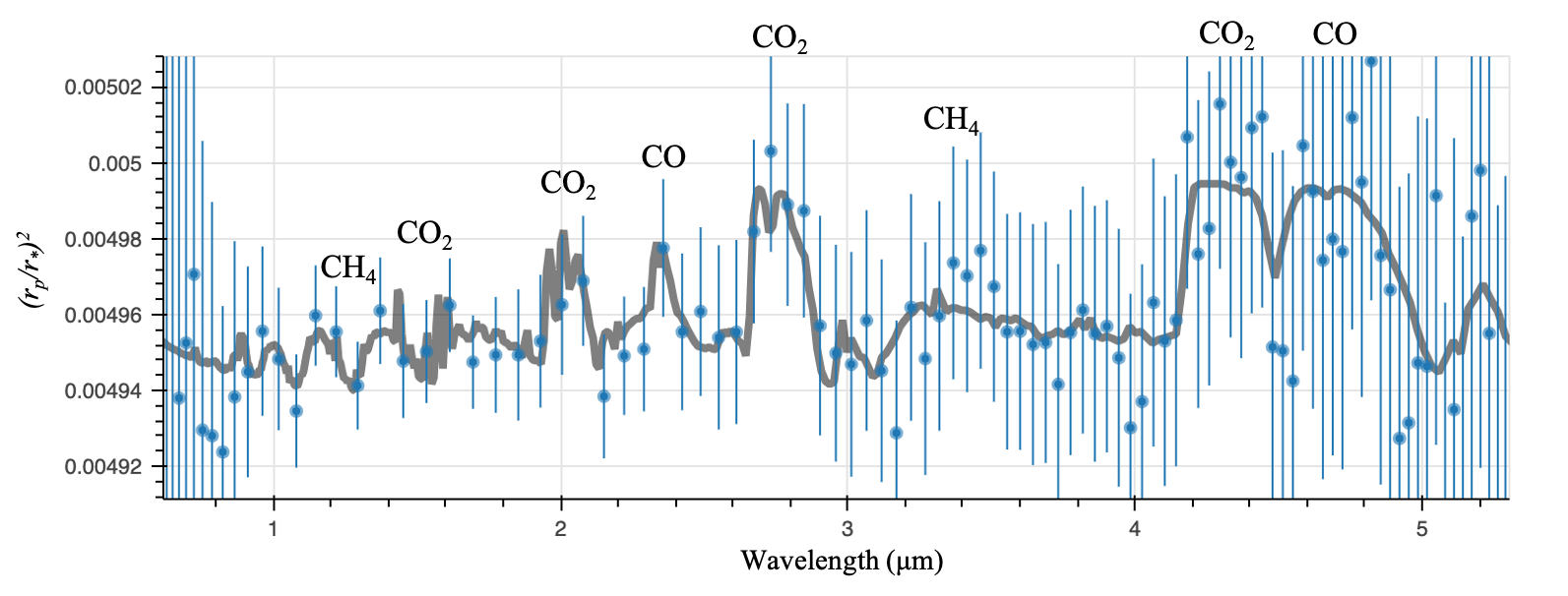}
    \caption{Synthetic observations generated using PandExo \citep{Batalha2017} of TRAPPIST-1e with Ecosystem 1 and $\Phi_{\mathrm{outgas}}(\ch{H2})$=100\,Tmol/yr. The observations are for NIRSpec with the Prism mode assume 20 transits, with four pixel bins. The transmission spectra used to generate the synthetic observations come from the PSG simulations presented in Figure~\ref{fig:spectra}. Visualisation of the data comes from \url{https://exoctk.stsci.edu/pandexo/}.
    \label{fig:pandexo}
    }
\end{figure*}

Although a \ch{CH4} feature is present in our model transmission spectrum it is unlikely to be detectable through observations using the JWST. Figure~\ref{fig:pandexo} shows the synthetic spectra for NIRSpec PRISM in the 0.8-5.1\,$\mu$m range for 20 transits (an effective upper limit) using PandExo \citep{Batalha2017}. Figure~\ref{fig:pandexo} shows that the \ch{CH4} features would not be robustly detected at 1.2, 1.4 and 3.3\,$\mu$m. \ch{CO2} at 2.7\,$\mu$m has the best change of detection, while \ch{CO} at 2.3\,$\mu$m could provide the best possibility for observation of this species. Beyond 5\,$\mu$m, using MIRI, there are no chances of detection in this range (not shown). Therefore, detection of these signatures might require next generation observational facilities such as the ELT.

\section{Discussion}
\label{sec:discussion}

From our simulations, biospheres that consume \ch{H2} and \ch{CO} producing \ch{CH4} as a waste product have the potential to be observed. Although detection of these may not be possible with JWST, future telescopes such as the ELT are likely to provide better opportunities for biosignature detection. Further, these results were specifically for TRAPPIST-1e, and other M-dwarf orbiting planets may prove to be better targets for observation. Signatures for \ch{CH4} are present when the reductant input is greater than 10\,Tmol/yr, which were absent in abiotic cases (see Figure~\ref{fig:spectra}). As well as this, strong features of \ch{CO2} as well as signs of water vapour are seen in all cases. This could be sufficient to suggest a biosphere with \citet{Krissansen-Totton2018earth-biosig} finding that this mix could provide a chemical disequilibrium strong enough to suggest the presence of life. However, strong \ch{CO} features in transmission spectra, despite the presence of \ch{CO} consuming organisms, may lead to some ambiguity as to whether this would be a definitively indicative biosphere \citep{Thompson2022}, despite near-surface concentrations of \ch{CO} being orders of magnitude smaller than \ch{CH4} (see Figure~\ref{fig:sun_trappist_comp}a). Our results do support the conclusions from \citet{Schwieterman2019} that \ch{CO} does not necessarily represent an anti-biosignature and can accumulate to high concentrations on inhabited planets, particularly for M--dwarf orbiting planets.

As our transmission spectra is generated using a 1D model only, without treatments of clouds and hazes, the effects of these have not been considered in the detectability. The presence of cloud may affect transmission spectra as discussed in \citet{Fauchez2019}, which may mask some features around 20ppm. This may mean that if TRAPPIST-1e had a lower reductant input, the detection of \ch{CH4} may not be possible, while a flux of greater than 1\,Tmol/yr would provide stronger signals. Similarly, a thin haze layer may be present in the highest H2 outgassing rate, which may clear many features below 3\,$\mu$m \citep{Fauchez2019}. However features of \ch{CH4} beyond 3\,$\mu$m may be detectable, but only when \ch{H2} outgassing is high, $\Phi_{\mathrm{outgas}}(\ch{H2})>$ 10\,Tmol/yr. We also find weaker \ch{CH4} peaks compared to \citet{Fauchez2019}. This may be because we find that \ch{CH4} photolysis peaks deeper in the atmosphere, meaning \ch{CH4} drops off substantially from around 50\,km compared to above 80\,km in \citet{Fauchez2019}. 

\citet{Segura2005} predicted that \ch{CH4} concentrations would be higher if the biological \ch{CH4} flux was the same as on Earth due to a lower rate of photochemical destruction. In addition to this, we find that the biological \ch{CH4} flux is generally lower on TRAPPIST-1e compared with the Earth due to the difference in stellar spectrum. TRAPPIST-1 has a lower proportion of flux in the NUV, which leads to a lower rate of \ch{H2O} photolysis which reduces the amount of tropospheric \ch{OH}. This is the predominant pathway for\ch{CH4} destruction below approximately 50\,km. The lower photochemical destruction of \ch{CH4} reduces the rate at which \ch{CO} and \ch{H2} are produced and leads to a lower biological methane flux, when the \ch{H2} outgassing is less than 10\,Tmol/yr. As a result we find that \ch{CH4} has a similar mixing ratio for both Earth and TRAPPIST-1e, with TRAPPIST-1e having only up to two times more \ch{CH4}. This difference is largest at a low reductant input, when there are high mixing ratios of oxygen, where the \ch{CH4} lifetime is increased further as the production of the \ch{OH} radical from \ch{H2O} photolysis is suppressed by an ozone layer.

\begin{figure}
    \centering
    \includegraphics[width=\linewidth]{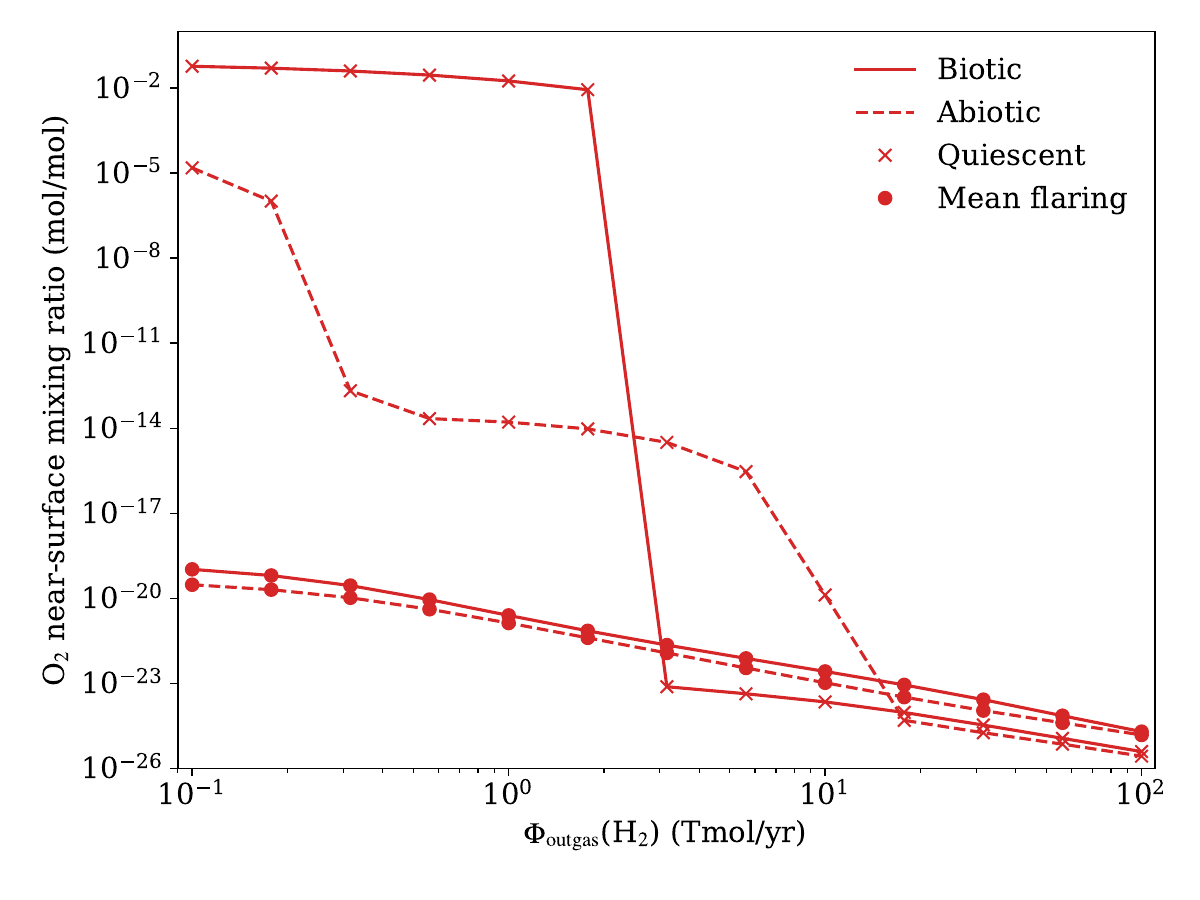}
    \caption{Near-surface (250\,m above the surface) \ch{O2} mixing ratio, for biotic and abiotic configurations for a quiescent and mean flaring spectrum of TRAPPIST-1.
    \label{fig:flaring}
    }
\end{figure}

Flaring has the potential to prevent the \ch{O2} build up found at low $\Phi_{\mathrm{outgas}}(\ch{H2})$ in the quiescent state by increasing the UV flux. We provide a preliminary exploration of this
using a mean flaring spectrum, shown in Figure~\ref{Fig:TOASpectra}, with the near-surface mixing ratios for key gases shown in Figure~\ref{fig:flaring}. The change in spectrum due to flaring could be sufficient to prevent \ch{O2} accumulating to values of greater than 1\%, seen in the quiescent state. The increased FUV flux due to flaring prevents \ch{O2} and importantly \ch{O3} accumulation, which maintains a weakly reducing atmosphere. This is by no means a comprehensive study of the effect of flaring, which would require time dependent modelling of atmospheric chemistry \citep[e.g.][]{Wogan2022,Ridgway2022}, but does provide a motivation for further work on the effects of flares.

An accurate vertical temperature profile is important to assess the potential for \ch{O2} accumulation. For M-dwarfs, the stratospheric temperature is sensitive to the \ch{CO2} and \ch{CH4} mixing ratio, due to the high proportion of infrared stellar radiation, with stratospheric temperatures of 190-250\,K plausible for TRAPPIST-1e \citep{Mak2024}. This can greatly affect the likelihood of atmospheric oxygen accumulation, and thus it is important to use temperature profiles that are more realistic for the atmospheric composition. This can be achieved through coupled climate chemistry models or using general circulation output in 1D chemistry models. The formation of ozone is unlikely to have a significant effect on the stratospheric temperature, as the ozone heating rate is much smaller for an M-dwarf spectrum \citep{Boutle2017,Yates2020,Kozakis2022}. This means it is unlikely that \ch{O3} formation could warm the stratosphere enough to then remove the ozone and produce either a stabilising negative feedback or an oscillating behaviour between a high and low ozone state. However, this could be investigated further in a coupled model.

The generation of high \ch{O2} states by anoxygenic biospheres is likely to be affected by the \ch{CO2} concentration. While we use a relatively low \ch{CO2} abundance of 10\%, higher \ch{CO2} levels could see high levels of \ch{O2} persisting for higher reductant inputs. 

The oxidising regime that emerges at these low \ch{H2} input levels in ecosystem 1, with only anoxic \ch{H2} and \ch{CO} consumers may also be missing important processes. At these levels of 10\%, processes such as oxidative weathering are likely to be important \citep{Daines2017}. More work is required to establish the plausibility of the states with high \ch{O2} levels found here due to different surface compositions.

The configuration used here effectively assumes a well-mixed rapidly-rotating planet. These planets are likely to be tidally-locked however, and either increasing the spatial resolution of 1D photochemical models to include day, night and terminator regions with appropriate mixing parametrisations or coupled chemistry-GCM are required to better constrain the detectability of these biospheres.

\section{Conclusion}

We use a coupled biosphere-atmosphere model to investigate the potential biosignatures from an \ch{H2}- and \ch{CO}-consuming biosphere, which produces \ch{CH4} as a waste product. We find that TRAPPIST-1e under a low reductant input with a \ch{CH4} producing biosphere may produce relatively high levels of near-surface \ch{O2} around 0.01--0.05\,mol/mol, which is not seen in the Earth case. This is primarily due to the consumption of \ch{CO} removing the major pathway for oxygen destruction in the atmosphere. In this high \ch{O2} scenario, \ch{O3} may be detectable at 9.4 microns, although this is difficult due to the presence of \ch{CO2}. The inclusion of \ch{O2} consuming organisms, could reduce these \ch{O2} levels to around 10$^{-5}$\,mol/mol. At higher reductant inputs, oxygen abundance decreases and strong features of \ch{CH4} along with \ch{CO2} may provide a robust biosignature. We also show that \ch{CO} is likely to be present in high quantities, even with \ch{CO} consuming organisms. The use of a mean flaring profile removes the high \ch{O2} states shown, however further work is required to understand how a time-dependent flaring spectrum may affect the results observed. We have shown that anoxigenic photosynthesisers are unlikely to affect the atmospheric composition significantly compared to a methanogenic biosphere, unless the recycling efficiency of secondary producers is low, where atmospheric \ch{CH4} becomes very low. These results show that simple biosphere models with consistent ocean fluxes are important in determining atmospheric composition for understanding the potential role of life on the atmospheric composition of exoplanets. The high oxygen state is sensitive to factors such as stratospheric temperature and thus further work using a coupled climate chemistry model are important for planets orbiting M-dwarfs.

\section*{Acknowledgements}
JKE-N would like to thank the Hill Family Scholarship. The Hill Family Scholarship has been generously supported by University of Exeter alumnus, and president of the University's US Foundation Graham Hill (Economic \& Political Development, 1992) and other donors to the US Foundation. Material produced using Met Office Software. NJM and TML gratefully acknowledge funding from a Leverhulme Trust Research Project Grant [RPG-2020-82]. SD and TL would to thank the John Templeton Foundation Grant [62220]. MTM acknowledge funding from the Bell Burnell Graduate Scholarship Fund, administered and managed by the Institute of Physics. This work was partly supported by a Science and Technology Facilities Council Consolidated Grant [ST/R000395/1]. This work was also supported by a UKRI Future Leaders Fellowship [grant number MR/T040866/1]. For the purpose of open access, the author(s) has applied a Creative Commons Attribution (CC BY) licence to any Author Accepted Manuscript version arising. \textbf{Author Contribution Statement:} JKE-N led this work, performing many of the model developments, conducting the final experiments, analysing the output and writing the manuscript. SJD led development of the PALEO framework, which formed the foundation for this work, as well as directly contributing to the scientific direction and model development of this work. JWMcD, PA, LAG, JB, AAR, JWGS, CK and TJB all contributed to model development, testing and scientific direction as part of undergraduate projects supervised by SJD, NJM and JKE-N. MTM, RJR, EH, FHL, TML \& NJM all contributed to scientific discussion of the work, and provided comments on the final manuscript. This work forms part of the PhD thesis of JKE-N, supervised by NJM \& TML. 

\section*{Data Availability}

The model output used for this study will be made available following this work's acceptance for publication.



\bibliographystyle{mnras}
\bibliography{references} 




\appendix

\section{Photochemical model}
\label{ref:network}

The Species list along with the deposition velocities employed here are in Table~\ref{tab:boundary_conditions}. Photolysis reactions are listed in Table~\ref{tab:photolysis}. For bimolecular reactions the rate coefficient, $k$, is calculated from the Arrhenius equation

\begin{equation}
    k = A\left(\frac{T}{300}\right)^\alpha e^{-\beta/T},
    \label{equ:bimolrate}
\end{equation}

where $A$ is a pre-exponential factor, $T$ is temperature in kelvins, $\alpha$ determines the degree of the temperature dependence for the reaction, $\beta$ is the divided by the gas constant. The parameters for each of the bimolecular reactions and their sources is listed in Table~\ref{tab:2body}.

For termolecular reactions, the rate coefficient is

\begin{equation}
    k = \frac{k_0[M]}{1+\frac{k_0}{k_{\infty}}[M]}F.
\end{equation}

\noindent where $k_0$ and $k_{\infty}$ are the low and high-pressure limits of the rate coefficient respectively:
    
\begin{equation}
    k_0 = A_1 \left(\frac{T}{300}\right)^{\alpha_1} e^{-\beta_1/T}
\end{equation}

\begin{equation}
    k_{\infty} = A_2 \left(\frac{T}{300}\right)^{\alpha_2} e^{-\beta_2/T}
\end{equation}

\noindent with the terms for these mirroring those of Equation~\ref{equ:bimolrate}. $F$ is a broadening factor, which takes the form
\begin{equation}
    F = F_c^{\frac{1}{1+(\log{P_r})^2}}
\end{equation}

\noindent where $F_c$ is a parameter greater than zero or less than or equal to one, and $P_r$, known as the reduced pressure is

\begin{equation}
    P_r = \frac{k_0}{k_{\infty}}[\ch{M}].
\end{equation}

\noindent The parameters for each of the termolecular reactions and their sources is listed in Table~\ref{tab:3body}.

\begin{table}
\begin{center}
 \caption{Photolysis reactions used in this network, with reactants (R) and products (P),
 with cross section and quantum yields from the open access ATMOS model \citep{Teal2022}. The horizontal line within the table shows the cutoff for the CHO network.}
 \label{tab:photolysis}
\begin{tabular}{lllll}
  \hline\textbf{R1} & \textbf{R2} &   \textbf{P1} &   \textbf{P2} & \textbf{P3} \\ \hline 
  \input{photolysis_reactions}
\hline
 \end{tabular}
 \end{center}
\end{table}

\begin{figure}
    \centering
    \includegraphics[width=\linewidth]{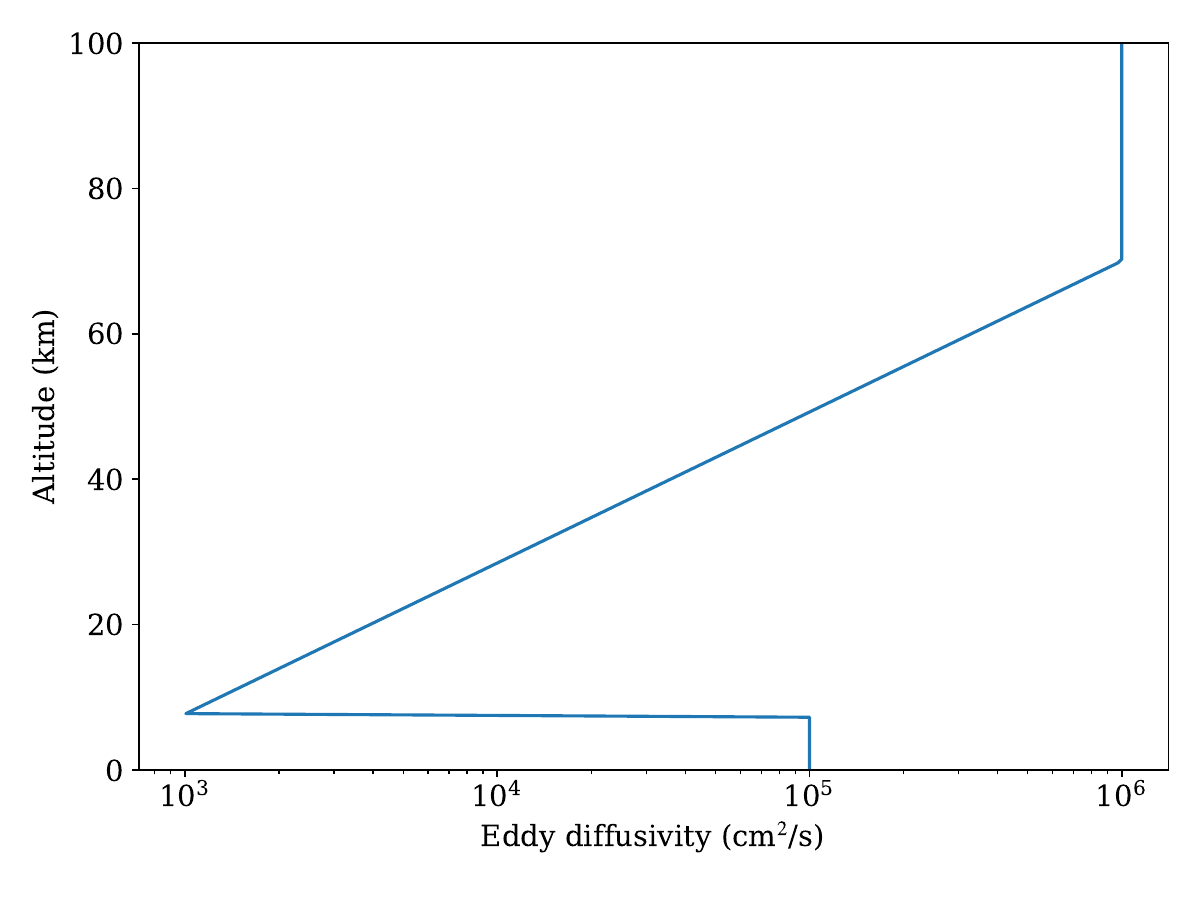}
    \caption{Eddy diffusion coefficient profile used in this study.
    \label{Fig:eddydiff}
    }
\end{figure}

The Eddy diffusion coefficient used in this study is shown in Figure~\ref{Fig:eddydiff}.

\onecolumn

\begin{center}
\begin{longtable}{llllllllll}
\caption{Bimolecular reactions used in this network. Reactions sourced from the Kinetic Database for Astrochemistry (KIDA) are from \url{https://kida.astrochem-tools.org/}. The horizontal line within the table shows the cutoff for the CHO network.} \label{tab:2body} \\

\hline\textbf{R1} & \textbf{R2} & \textbf{R3} &   \textbf{P1} &   \textbf{P2} & \textbf{P3} & \textbf{$\bm{A}$} & \textbf{$\bm{\alpha}$} & \textbf{$\bm{\beta}$ (K)} &  \textbf{Source} \\ \hline 
\endfirsthead

\multicolumn{10}{c}%
{{\bfseries \tablename\ \thetable{} -- continued from previous page}} \\
\hline \textbf{R1} & \textbf{R2} & \textbf{R3} &   \textbf{P1} &   \textbf{P2} & \textbf{P3} & \textbf{$\bm{A}$} & \textbf{$\bm{\alpha}$} & \textbf{$\bm{\beta}$ (K)} &  \textbf{Source} \\ \hline 
\endhead

\multicolumn{10}{r}{{\textbf{Continued on next page}}} \\ 

\endfoot

\hline \hline
\endlastfoot

\input{2body_reactions}
\end{longtable}
\end{center}

\begin{center}
\small
\begin{longtable}{llllllllllll}

\caption[Termolecular reaction table]{Termolecular reactions used in this network. Reactions sourced from the Kinetic Database for Astrochemistry (KIDA) are from \url{https://kida.astrochem-tools.org/} and those sourced from the International Union of Pure and Applied Chemistry (IUPAC) are obtained from \url{https://iupac-aeris.ipsl.fr/}. The horizontal line within the table shows the cutoff for the CHO network.} \label{tab:3body} \\

\hline\textbf{R1} & \textbf{R2} &   \textbf{P1} &   \textbf{P2} & \textbf{$\bm{A_1}$} & \textbf{$\bm{\alpha_1}$} & \textbf{$\bm{\beta_1}$ (K)} & \textbf{$\bm{A_2}$} & \textbf{$\bm{\alpha_2}$} & \textbf{$\bm{\beta_2}$ (K)} & \textbf{$\bm{F_c}$} &  \textbf{Source} \\ \hline 
\endfirsthead

\multicolumn{12}{c}%
{{\bfseries \tablename\ \thetable{} -- continued from previous page}} \\
\hline\textbf{R1} & \textbf{R2} &   \textbf{P1} &   \textbf{P2} & \textbf{$\bm{A_1}$} & \textbf{$\bm{\alpha_1}$} & \textbf{$\bm{\beta_1}$ (K)} & \textbf{$\bm{A_2}$} & \textbf{$\bm{\alpha_2}$} & \textbf{$\bm{\beta_2}$ (K)} & \textbf{$\bm{F_c}$} &  \textbf{Source} \\ \hline 
\endhead

\multicolumn{12}{r}{{\textbf{Continued on next page}}} \\ 

\endfoot

\hline \hline
\endlastfoot

\input{3body_reactions}
\end{longtable}
\end{center}

\bsp	
\label{lastpage}
\end{document}

%% file: species_list.tex
\ch{CO2} & & & & 0.1  & &               \ch{SO2} & & 0.802   & 2.97$\times10^{9}$ & & 1.0 \\
\ch{N2} & & & & 0.88   & &                  \ch{H2S} & & 0.0802 & 2.97$\times10^{8}$ &   & \\
\ch{H2O} &   & & &   &  &               NO & & 0.16 & 5.93$\times10^{8}$ &  & \\
\ch{CH4} & & 0.0802  & 2.97$\times10^{8}$ &  & &               \ch{CS2} & &  & &   & \\
\ch{H2} & & 0.1--100  & 3.70--3700$\times10^{8}$ & & &                 \ch{CS2^*} & y  & & &  & \\
CO &   & 0.16$^{\dag}$ & 5.93$\times10^{8}$ &  & &                     HS & &  &  &   & \\
\ch{O3} & & &   &  & 0.4 &           CS &  & &   & & 0.01 \\
\ch{O2} &  & &  &  &     &          \ch{HNO2}   & y  & & &  & \\
\ch{HO2} &  & & & & 1.0 &            \ch{HNO3} &  &  & &  & 0.2 \\
\ch{H2O2} &  & & & & 0.5 &           \ch{NO2} & &   &  & & 3$\times10^{-3}$\\
O($^3$P) &  & & & & 1.0 &            N  &  &  & &  & \\
O($^1$D) & y & & & &  &              HNO &  &  & &  & 1.0 \\
\ch{OH} &  & & & & 1.0 &             \ch{NO3} &  & & &   & \\
HCO &  & & & & 0.1 &                 SO &  & & &   & \\
\ch{H2CO} & & &  & & 0.1 &           S & &  & &   & \\
H &  & & & & 1.0 &                   \ch{^1SO2}  & y  & &  & & \\
\ch{CH3} & & & &  &  &               \ch{^3SO2}  & y  & & &  & \\
\ch{^1CH2} & y & &  & &  &           \ch{HSO3}  & y & &  & & \\
\ch{^3CH2} &  & & & & &              \ch{SO3} & &  & &   & \\
\ch{CH3O} & & &  & & 0.1 &           \ch{H2SO4} & &  & &   & \\
\ch{C2H6} & & & &  & 1$\times10^{-5}$ & HSO &  &  & &  & 1.0 \\
\ch{C2H5} & & & &  & &               \ch{S2} &  & &  &  & \\
\ch{C2H4} & & & &  & &               OCS &  &  & &  & 0.01 \\
\ch{CH3O2} &  & & & & &              \ch{S3} &  & &  &  & \\
\ch{C2H2} & & & &  & &               \ch{S4} &  & &  &  & \\
\ch{CH3CHO} & & & &  & 0.1 &         \ch{S8} &  &  & &  & \\
C &  & &  & &  &                     \ch{OCS2}  & y &  & &  & \\
CH &  & &  & &  &                    \ch{N2O5} &  & & \\
\ch{CH2CO} & & & &  & 0.1 &          \ch{HO2NO2} &  &  & &  & 0.2 \\
\ch{CH3CO} & & & &  & 0.1 &          \ch{N2O} &  & &  &  & \\
\ch{C2H} & & & &  & &                HCS &  &  & &  & \\
\ch{C2} & & & &  & &                 &   & &   & \\
\ch{C2H3} & & &  & & &               &   & &   & \\
\ch{C2H2OH} & & & &  & &             &   & &   & \\
\ch{C2H4OH} & & &  & & &             &   & &   & \\

%% file: photolysis_reactions.tex
    \ch{O2}& $h\nu$ &    O($^3$P)&  O($^1$D)&  - \\
    \ch{O2}& $h\nu$ &    O($^3$P)&    O($^3$P)&  -  \\
   \ch{H2O}& $h\nu$ &    H &   OH &  -  \\
   \ch{H2O}& $h\nu$ &   \ch{H2}&  O($^1$D)&  -                    \\
   \ch{H2O}& $h\nu$ &    H &    H &  O($^3$P)                  \\
    \ch{O3}& $h\nu$ &   \ch{O2}&  O($^1$D)&  -  \\
    \ch{O3}& $h\nu$ &   \ch{O2}&    O($^3$P)&  -  \\

%% file: 2body_reactions.tex
      O($^1$D)   &      \ch{H2O}  &         -  &            OH  &           OH  &        -  & 1.63e-10 &  0.000 &   -60 &                                             \citet{Burkholder2019} \\
      O($^1$D)   &       \ch{H2}  &         -  &            OH  &            H  &        -  & 1.20e-10 &  0.000 &     0 &                                             \citet{Burkholder2019} \\
     \ch{H2}  &             O($^3$P) &         -  &            OH  &            H  &        -  & 3.44e-13 &  2.670 &  3160 &                               \citet{Baulch1992} \\
          OH  &       \ch{H2}  &         -  &      \ch{H2O}  &            H  &        -  & 2.80e-12 &  0.000 &  1800 &                                             \citet{Burkholder2019} \\
           H  &       \ch{O3}  &         -  &            OH  &      \ch{O2}  &        -  & 1.40e-10 &  0.000 &   470 &                                             \citet{Burkholder2019} \\
           H  &      \ch{HO2}  &         -  &       \ch{H2}  &      \ch{O2}  &        -  & 6.90e-12 &  0.000 &     0 &                                             \citet{Burkholder2019} \\
           H  &      \ch{HO2}  &         -  &             O($^3$P) &     \ch{H2O}  &        -  & 1.60e-12 &  0.000 &     0 &                                             \citet{Burkholder2019} \\
           H  &      \ch{HO2}  &         -  &            OH  &           OH  &        -  & 7.20e-11 &  0.000 &     0 &                                             \citet{Burkholder2019} \\

%% file: 3body_reactions.tex
     H &  \ch{O2} &   \ch{HO2} &   - & 5.30e-32 & -1.800 &     0 & 9.50e-11 &  0.400 &     0 & 0.600 &                                             \citet{Burkholder2019} \\
  \ch{HO2}& \ch{HO2}&  \ch{H2O2}& \ch{O2}& 2.10e-33 &  0.000 &  -920 & 1.00e-10 &  0.000 &     0 & 0.600 &                 \citet{Burkholder2019} \\
     O(3P)&    O(3P)&    \ch{O2}&   - & 9.46e-34 &  0.000 &  -485 & 1.00e-10 &  0.000 &     0 & 0.600 & \citet{CampbellGray1973} \\
     O(3P)&  \ch{O2}&    \ch{O3}&   - & 6.10e-34 & -2.400 &     0 & 1.00e-10 &  0.000 &     0 & 0.600 &                 \citet{Burkholder2019} \\
    CO &    O(3P)&   \ch{CO2}&   - & 6.51e-33 &  0.000 &  2188 & 1.00e-10 &  0.000 &     0 & 0.600 &  \citet{Slanger1972} \\
     H &   CO &    HCO &   - & 5.29e-34 &  0.000 &   373 & 1.00e-10 &  0.000 &     0 & 0.600 &                                               KIDA \\
     H &    H &    \ch{H2}&   - & 8.85e-33 & -0.600 &     0 & 1.00e-10 &  0.000 &     0 & 0.600 &   \citet{Baulch1992} \\
     H &   OH &   \ch{H2O}&   - & 6.78e-31 & -2.000 &     0 & 1.00e-10 &  0.000 &     0 & 0.600 &                                               KIDA \\

%% file: mnras_template.bbl
\begin{thebibliography}{}
\makeatletter
\relax
\def\mn@urlcharsother{\let\do\@makeother \do\$\do\&\do\#\do\^\do\_\do\%\do\~}
\def\mn@doi{\begingroup\mn@urlcharsother \@ifnextchar [ {\mn@doi@} {\mn@doi@[]}}
\def\mn@doi@[#1]#2{\def\@tempa{#1}\ifx\@tempa\@empty \href {http://dx.doi.org/#2} {doi:#2}\else \href {http://dx.doi.org/#2} {#1}\fi \endgroup}
\def\mn@eprint#1#2{\mn@eprint@#1:#2::\@nil}
\def\mn@eprint@arXiv#1{\href {http://arxiv.org/abs/#1} {{\tt arXiv:#1}}}
\def\mn@eprint@dblp#1{\href {http://dblp.uni-trier.de/rec/bibtex/#1.xml} {dblp:#1}}
\def\mn@eprint@#1:#2:#3:#4\@nil{\def\@tempa {#1}\def\@tempb {#2}\def\@tempc {#3}\ifx \@tempc \@empty \let \@tempc \@tempb \let \@tempb \@tempa \fi \ifx \@tempb \@empty \def\@tempb {arXiv}\fi \@ifundefined {mn@eprint@\@tempb}{\@tempb:\@tempc}{\expandafter \expandafter \csname mn@eprint@\@tempb\endcsname \expandafter{\@tempc}}}

\bibitem[\protect\citeauthoryear{Adachi, Basco  \& James}{Adachi et~al.}{1981}]{Adachi1981}
Adachi H.,  Basco N.,   James D. G.~L.,  1981, \mn@doi [International Journal of Chemical Kinetics] {10.1002/kin.550131206}, 13, 1251

\bibitem[\protect\citeauthoryear{Agol et~al.,}{Agol et~al.}{2021}]{Agol2021}
Agol E.,  et~al., 2021, \mn@doi [The Planetary Science Journal] {10.3847/PSJ/abd022}, 2, 1

\bibitem[\protect\citeauthoryear{Arthur, Dean, Neff, Hay, King  \& Jones}{Arthur et~al.}{1994}]{Arthur1994}
Arthur M.~A.,  Dean W.~E.,  Neff E.~D.,  Hay B.~J.,  King J.,   Jones G.,  1994, \mn@doi [Global Biogeochemical Cycles] {10.1029/94GB00297}, 8, 195

\bibitem[\protect\citeauthoryear{Babikov, Semenov  \& Teplukhin}{Babikov et~al.}{2017}]{Babikov2017}
Babikov D.,  Semenov A.,   Teplukhin A.,  2017, \mn@doi [Geochimica et Cosmochimica Acta] {10.1016/j.gca.2017.01.029}, 204, 388

\bibitem[\protect\citeauthoryear{Baggott, Frey, Lightfoot  \& Walsh}{Baggott et~al.}{1987}]{Baggott1987}
Baggott J.~E.,  Frey H.~M.,  Lightfoot P.~D.,   Walsh R.,  1987, \mn@doi [The Journal of Physical Chemistry] {10.1021/j100296a057}, 91, 3386

\bibitem[\protect\citeauthoryear{Basco \& Pearson}{Basco \& Pearson}{1967}]{BascoPearson1967}
Basco N.,  Pearson A.~E.,  1967, \mn@doi [Transactions of the Faraday Society] {10.1039/tf9676302684}, 63, 2684

\bibitem[\protect\citeauthoryear{Batalha et~al.,}{Batalha et~al.}{2017}]{Batalha2017}
Batalha N.~E.,  et~al., 2017, \mn@doi [Publications of the Astronomical Society of the Pacific] {10.1088/1538-3873/aa65b0}, 129, 064501

\bibitem[\protect\citeauthoryear{Battistuzzi, Feijao  \& Hedges}{Battistuzzi et~al.}{2004}]{Battistuzzi2004}
Battistuzzi F.~U.,  Feijao A.,   Hedges S.~B.,  2004, \mn@doi [BMC Evolutionary Biology] {10.1186/1471-2148-4-44}, 4, 44

\bibitem[\protect\citeauthoryear{Baulch et~al.,}{Baulch et~al.}{1992}]{Baulch1992}
Baulch D.~L.,  et~al., 1992, \mn@doi [Journal of Physical and Chemical Reference Data] {10.1063/1.555908}, 21, 411

\bibitem[\protect\citeauthoryear{Baulch et~al.,}{Baulch et~al.}{1994}]{Baulch1994}
Baulch D.~L.,  et~al., 1994, \mn@doi [Journal of Physical and Chemical Reference Data] {10.1063/1.555953}, 23, 847

\bibitem[\protect\citeauthoryear{Baulch et~al.,}{Baulch et~al.}{2005}]{Baulch2005}
Baulch D.~L.,  et~al., 2005, \mn@doi [Journal of Physical and Chemical Reference Data] {10.1063/1.1748524}, 34, 757

\bibitem[\protect\citeauthoryear{Berner}{Berner}{1982}]{Berner1982}
Berner R.~A.,  1982, \mn@doi [American Journal of Science] {10.2475/ajs.282.4.451}, 282, 451

\bibitem[\protect\citeauthoryear{Boutle, Mayne, Drummond, Manners, Goyal, Hugo~Lambert, Acreman  \& Earnshaw}{Boutle et~al.}{2017}]{Boutle2017}
Boutle I.~A.,  Mayne N.~J.,  Drummond B.,  Manners J.,  Goyal J.,  Hugo~Lambert F.,  Acreman D.~M.,   Earnshaw P.~D.,  2017, \mn@doi [Astronomy {\&} Astrophysics] {10.1051/0004-6361/201630020}, 601, A120

\bibitem[\protect\citeauthoryear{Brune, Schwab  \& Anderson}{Brune et~al.}{1983}]{Brune1983}
Brune W.~H.,  Schwab J.~J.,   Anderson J.~G.,  1983, \mn@doi [The Journal of Physical Chemistry] {10.1021/j100245a034}, 87, 4503

\bibitem[\protect\citeauthoryear{Burkholder et~al.,}{Burkholder et~al.}{2015}]{Burkholder2015JPL}
Burkholder J.~B.,  et~al., 2015, Technical report, {"Chemical Kinetics and Photochemical Data for Use in Atmospheric Studies, Evaluation No. 18," JPL Publication 15-10}, \url {http://jpldataeval.jpl.nasa.gov}.
Jet Propulsion Laboratory, Pasadena, \url {http://jpldataeval.jpl.nasa.gov}

\bibitem[\protect\citeauthoryear{Burkholder et~al.,}{Burkholder et~al.}{2019}]{Burkholder2019}
Burkholder J.~B.,  et~al., 2019, Technical report, {Chemical Kinetics and Photochemical Data for Use in Atmospheric Studies, Evaluation No. 19,}.
Jet Propulsion Laboratory, Pasadena

\bibitem[\protect\citeauthoryear{Campbell \& Gray}{Campbell \& Gray}{1973}]{CampbellGray1973}
Campbell I.,  Gray C.,  1973, \mn@doi [Chemical Physics Letters] {10.1016/0009-2614(73)80479-8}, 18, 607

\bibitem[\protect\citeauthoryear{Canosa-mas, Frey  \& Walsh}{Canosa-mas et~al.}{1984}]{Canosa-Mas1984}
Canosa-mas C.~E.,  Frey H.~M.,   Walsh R.,  1984, \mn@doi [J. Chem. Soc., Faraday Trans. 2] {10.1039/F29848000561}, 80, 561

\bibitem[\protect\citeauthoryear{Catling et~al.,}{Catling et~al.}{2018}]{Catling2018}
Catling D.~C.,  et~al., 2018, \mn@doi [Astrobiology] {10.1089/ast.2017.1737}, 18, 709

\bibitem[\protect\citeauthoryear{Chen, Zhan, Youngblood, Wolf, Feinstein  \& Horton}{Chen et~al.}{2020}]{Chen2020}
Chen H.,  Zhan Z.,  Youngblood A.,  Wolf E.~T.,  Feinstein A.~D.,   Horton D.~E.,  2020, \mn@doi [Nature Astronomy] {10.1038/s41550-020-01264-1}, 5, 298

\bibitem[\protect\citeauthoryear{Choi \& Lin}{Choi \& Lin}{2005}]{ChoiLin2005}
Choi Y.~M.,  Lin M.~C.,  2005, \mn@doi [International Journal of Chemical Kinetics] {10.1002/kin.20079}, 37, 261

\bibitem[\protect\citeauthoryear{Chung, Calvert  \& Bottenheim}{Chung et~al.}{1975}]{Chung1975}
Chung K.,  Calvert J.~G.,   Bottenheim J.~W.,  1975, \mn@doi [International Journal of Chemical Kinetics] {10.1002/kin.550070202}, 7, 161

\bibitem[\protect\citeauthoryear{Claire, Sheets, Cohen, Ribas, Meadows  \& Catling}{Claire et~al.}{2012}]{Claire2012}
Claire M.~W.,  Sheets J.,  Cohen M.,  Ribas I.,  Meadows V.~S.,   Catling D.~C.,  2012, \mn@doi [The Astrophysical Journal] {10.1088/0004-637X/757/1/95}, 757, 95

\bibitem[\protect\citeauthoryear{Daines \& Lenton}{Daines \& Lenton}{2016}]{DainesLenton2016}
Daines S.~J.,  Lenton T.~M.,  2016, \mn@doi [Earth and Planetary Science Letters] {10.1016/j.epsl.2015.11.021}, 434, 42

\bibitem[\protect\citeauthoryear{Daines, Mills  \& Lenton}{Daines et~al.}{2017}]{Daines2017}
Daines S.~J.,  Mills B. J.~W.,   Lenton T.~M.,  2017, \mn@doi [Nature Communications] {10.1038/ncomms14379}, 8, 14379

\bibitem[\protect\citeauthoryear{Dammeier, Colberg  \& Friedrichs}{Dammeier et~al.}{2007}]{Dammeier2007}
Dammeier J.,  Colberg M.,   Friedrichs G.,  2007, \mn@doi [Physical Chemistry Chemical Physics] {10.1039/b704197g}, 9, 4177

\bibitem[\protect\citeauthoryear{Davidson, Schiff, Brown  \& Howard}{Davidson et~al.}{1978}]{Davidson1978}
Davidson J.~A.,  Schiff H.~I.,  Brown T.~J.,   Howard C.~J.,  1978, \mn@doi [The Journal of Chemical Physics] {10.1063/1.436657}, 69, 1216

\bibitem[\protect\citeauthoryear{Davies, Duarte  \& Green}{Davies et~al.}{2023}]{Davies2023}
Davies H.~S.,  Duarte J.~C.,   Green M.,  2023, in , A Journey Through Tides.
Elsevier, pp 133--141, \mn@doi{10.1016/B978-0-323-90851-1.00020-0}

\bibitem[\protect\citeauthoryear{Devriendt \& Peeters}{Devriendt \& Peeters}{1997}]{DevriendtPeeters1997}
Devriendt K.,  Peeters J.,  1997, \mn@doi [The Journal of Physical Chemistry A] {10.1021/jp963434i}, 101, 2546

\bibitem[\protect\citeauthoryear{Dillon, Horowitz  \& Crowley}{Dillon et~al.}{2007}]{Dillon2007}
Dillon T.~J.,  Horowitz A.,   Crowley J.~N.,  2007, \mn@doi [Chemical Physics Letters] {10.1016/j.cplett.2007.06.044}, 443, 12

\bibitem[\protect\citeauthoryear{Domagal-Goldman, Meadows, Claire  \& Kasting}{Domagal-Goldman et~al.}{2011}]{Domagal-Goldman2011}
Domagal-Goldman S.~D.,  Meadows V.~S.,  Claire M.~W.,   Kasting J.~F.,  2011, \mn@doi [Astrobiology] {10.1089/ast.2010.0509}, 11, 419

\bibitem[\protect\citeauthoryear{Domagal-Goldman, Segura, Claire, Robinson  \& Meadows}{Domagal-Goldman et~al.}{2014}]{Domagal-Goldman2014}
Domagal-Goldman S.~D.,  Segura A.,  Claire M.~W.,  Robinson T.~D.,   Meadows V.~S.,  2014, \mn@doi [The Astrophysical Journal] {10.1088/0004-637X/792/2/90}, 792, 90

\bibitem[\protect\citeauthoryear{Du, Francisco, Shepler  \& Peterson}{Du et~al.}{2008}]{Du2008}
Du S.,  Francisco J.~S.,  Shepler B.~C.,   Peterson K.~A.,  2008, \mn@doi [The Journal of Chemical Physics] {10.1063/1.2919569}, 128

\bibitem[\protect\citeauthoryear{Du, Germann, Francisco, Peterson, Yu  \& Lyons}{Du et~al.}{2011}]{Du2011}
Du S.,  Germann T.~C.,  Francisco J.~S.,  Peterson K.~A.,  Yu H.-G.,   Lyons J.~R.,  2011, \mn@doi [The Journal of Chemical Physics] {10.1063/1.3572226}, 134

\bibitem[\protect\citeauthoryear{Eager-Nash et~al.,}{Eager-Nash et~al.}{2020}]{Eager-Nash2020}
Eager-Nash J.~K.,  et~al., 2020, \mn@doi [Astronomy {\&} Astrophysics] {10.1051/0004-6361/202038089}, 639, A99

\bibitem[\protect\citeauthoryear{Eager‐Nash et~al.,}{Eager‐Nash et~al.}{2023}]{Eager-Nash2023}
Eager‐Nash J.~K.,  et~al., 2023, \mn@doi [Journal of Geophysical Research: Atmospheres] {10.1029/2022JD037544}, 128

\bibitem[\protect\citeauthoryear{Edwards \& Slingo}{Edwards \& Slingo}{1996}]{EdwardsSlingo1996}
Edwards J.~M.,  Slingo A.,  1996, \mn@doi [Quarterly Journal of the Royal Meteorological Society] {10.1002/qj.49712253107}, 122, 689

\bibitem[\protect\citeauthoryear{Fardeau \& Belaich}{Fardeau \& Belaich}{1986}]{FardeauBelaich1986}
Fardeau M.-L.,  Belaich J.-P.,  1986, \mn@doi [Archives of Microbiology] {10.1007/BF00409888}, 144, 381

\bibitem[\protect\citeauthoryear{Fauchez et~al.,}{Fauchez et~al.}{2019}]{Fauchez2019}
Fauchez T.~J.,  et~al., 2019, \mn@doi [The Astrophysical Journal] {10.3847/1538-4357/ab5862}, 887, 194

\bibitem[\protect\citeauthoryear{Fauchez et~al.,}{Fauchez et~al.}{2020}]{Fauchez2020}
Fauchez T.,  et~al., 2020, \mn@doi [Geosci. Model Dev.] {10.5194/gmd-13-707-2020}, 13, 707

\bibitem[\protect\citeauthoryear{Ferry}{Ferry}{2006}]{Ferry2006}
Ferry J.~G.,  2006, \mn@doi [Molecular Biology and Evolution] {10.1093/molbev/msk014}, 23, 1286

\bibitem[\protect\citeauthoryear{Gao \& Marshall}{Gao \& Marshall}{2011}]{GaoMarshall2011}
Gao Y.,  Marshall P.,  2011, \mn@doi [The Journal of Chemical Physics] {10.1063/1.3644773}, 135

\bibitem[\protect\citeauthoryear{Gauthier \& Snelling}{Gauthier \& Snelling}{1975}]{GauthierSnelling1975}
Gauthier M.~J.,  Snelling D.~R.,  1975, \mn@doi [Journal of Photochemistry] {10.1016/0047-2670(75)80012-8}, 4, 27

\bibitem[\protect\citeauthoryear{Gillon et~al.,}{Gillon et~al.}{2017}]{Gillon2017}
Gillon M.,  et~al., 2017, \mn@doi [Nature] {10.1038/nature21360}, 542, 456

\bibitem[\protect\citeauthoryear{Giorgi \& Chameides}{Giorgi \& Chameides}{1985}]{GiorgiChameides1985}
Giorgi F.,  Chameides W.~L.,  1985, \mn@doi [Journal of Geophysical Research: Atmospheres] {10.1029/JD090iD05p07872}, 90, 7872

\bibitem[\protect\citeauthoryear{Gladstone, Allen  \& Yung}{Gladstone et~al.}{1996}]{Gladstone1996}
Gladstone G.,  Allen M.,   Yung Y.,  1996, \mn@doi [Icarus] {10.1006/icar.1996.0001}, 119, 1

\bibitem[\protect\citeauthoryear{Gough}{Gough}{1981}]{Gough1981}
Gough D.~O.,  1981, \mn@doi [Solar Physics] {10.1007/BF00151270}, 74, 21

\bibitem[\protect\citeauthoryear{Greene, Bell, Ducrot, Dyrek, Lagage  \& Fortney}{Greene et~al.}{2023}]{Greene2023}
Greene T.~P.,  Bell T.~J.,  Ducrot E.,  Dyrek A.,  Lagage P.-O.,   Fortney J.~J.,  2023, \mn@doi [Nature] {10.1038/s41586-023-05951-7}, 618, 39

\bibitem[\protect\citeauthoryear{Gregory, Claire  \& Rugheimer}{Gregory et~al.}{2021}]{Gregory2021}
Gregory B.~S.,  Claire M.~W.,   Rugheimer S.,  2021, \mn@doi [Earth and Planetary Science Letters] {https://doi.org/10.1016/j.epsl.2021.116818}, 561, 116818

\bibitem[\protect\citeauthoryear{Guimond, Noack, Ortenzi  \& Sohl}{Guimond et~al.}{2021}]{Guimond2021}
Guimond C.~M.,  Noack L.,  Ortenzi G.,   Sohl F.,  2021, \mn@doi [Physics of the Earth and Planetary Interiors] {10.1016/j.pepi.2021.106788}, 320, 106788

\bibitem[\protect\citeauthoryear{Harman, Schwieterman, Schottelkotte  \& Kasting}{Harman et~al.}{2015}]{Harman2015}
Harman C.~E.,  Schwieterman E.~W.,  Schottelkotte J.~C.,   Kasting J.~F.,  2015, \mn@doi [The Astrophysical Journal] {10.1088/0004-637X/812/2/137}, 812, 137

\bibitem[\protect\citeauthoryear{Harman, Felton, Hu, Domagal-Goldman, Segura, Tian  \& Kasting}{Harman et~al.}{2018}]{Harman2018}
Harman C.~E.,  Felton R.,  Hu R.,  Domagal-Goldman S.~D.,  Segura A.,  Tian F.,   Kasting J.~F.,  2018, \mn@doi [The Astrophysical Journal] {10.3847/1538-4357/aadd9b}, 866, 56

\bibitem[\protect\citeauthoryear{Harris \& B{\'{e}}dard}{Harris \& B{\'{e}}dard}{2014}]{HarrisBedard2014}
Harris L.~B.,  B{\'{e}}dard J.~H.,  2014, Springer, Dordrecht, pp 215--291, \mn@doi{10.1007/978-94-007-7615-9{\_}9}

\bibitem[\protect\citeauthoryear{Hassinen, Kalliorinne  \& Koskikallio}{Hassinen et~al.}{1990}]{Hassinen1990}
Hassinen E.,  Kalliorinne K.,   Koskikallio J.,  1990, \mn@doi [International Journal of Chemical Kinetics] {10.1002/kin.550220709}, 22, 741

\bibitem[\protect\citeauthoryear{Hawley, Davenport, Kowalski, Wisniewski, Hebb, Deitrick  \& Hilton}{Hawley et~al.}{2014}]{Hawley2014}
Hawley S.~L.,  Davenport J. R.~A.,  Kowalski A.~F.,  Wisniewski J.~P.,  Hebb L.,  Deitrick R.,   Hilton E.~J.,  2014, \mn@doi [The Astrophysical Journal] {10.1088/0004-637X/797/2/121}, 797, 121

\bibitem[\protect\citeauthoryear{Herron}{Herron}{1988}]{Herron1988}
Herron J.~T.,  1988, \mn@doi [Journal of Physical and Chemical Reference Data] {10.1063/1.555810}, 17, 967

\bibitem[\protect\citeauthoryear{Hu, Seager  \& Bains}{Hu et~al.}{2012}]{Hu2012}
Hu R.,  Seager S.,   Bains W.,  2012, \mn@doi [The Astrophysical Journal] {10.1088/0004-637X/761/2/166}, 761, 166

\bibitem[\protect\citeauthoryear{Hu, Peterson  \& Wolf}{Hu et~al.}{2020}]{Hu2020}
Hu R.,  Peterson L.,   Wolf E.~T.,  2020, \mn@doi [The Astrophysical Journal] {10.3847/1538-4357/ab5f07}, 888, 122

\bibitem[\protect\citeauthoryear{Ih, Kempton, Whittaker  \& Lessard}{Ih et~al.}{2023}]{Ih2023}
Ih J.,  Kempton E. M.-R.,  Whittaker E.~A.,   Lessard M.,  2023, \mn@doi [The Astrophysical Journal Letters] {10.3847/2041-8213/ace03b}, 952, L4

\bibitem[\protect\citeauthoryear{Karman et~al.,}{Karman et~al.}{2019}]{Karman2019}
Karman T.,  et~al., 2019, \mn@doi [Icarus] {10.1016/j.icarus.2019.02.034}, 328, 160

\bibitem[\protect\citeauthoryear{Kasting}{Kasting}{1990}]{Kasting1990}
Kasting J.~F.,  1990, \mn@doi [Origins of Life and Evolution of the Biosphere] {10.1007/BF01808105}, 20, 199

\bibitem[\protect\citeauthoryear{Kasting, Eggler  \& Raeburn}{Kasting et~al.}{1993}]{Kasting1993redox}
Kasting J.~F.,  Eggler D.~H.,   Raeburn S.~P.,  1993, \mn@doi [The Journal of Geology] {10.1086/648219}, 101, 245

\bibitem[\protect\citeauthoryear{Kasting, Pavlov  \& Siefert}{Kasting et~al.}{2001}]{Kasting2001}
Kasting J.~F.,  Pavlov A.~A.,   Siefert J.~L.,  2001, \mn@doi [Origins of Life and Evolution of the Biosphere] {10.1023/A:1010600401718}, 31, 271

\bibitem[\protect\citeauthoryear{Kharecha, Kasting  \& Siefert}{Kharecha et~al.}{2005}]{Kharecha2005}
Kharecha P.,  Kasting J.,   Siefert J.,  2005, \mn@doi [Geobiology] {10.1111/j.1472-4669.2005.00049.x}, 3, 53

\bibitem[\protect\citeauthoryear{Knoll \& Nowak}{Knoll \& Nowak}{2017}]{Knoll2017}
Knoll A.~H.,  Nowak M.~A.,  2017, \mn@doi [Science Advances] {10.1126/sciadv.1603076}, 3

\bibitem[\protect\citeauthoryear{Korenaga}{Korenaga}{2013}]{Korenaga2013}
Korenaga J.,  2013, \mn@doi [Annual Review of Earth and Planetary Sciences] {10.1146/annurev-earth-050212-124208}, 41, 117

\bibitem[\protect\citeauthoryear{Kozakis, Mendon{\c{c}}a  \& Buchhave}{Kozakis et~al.}{2022}]{Kozakis2022}
Kozakis T.,  Mendon{\c{c}}a J.~M.,   Buchhave L.~A.,  2022, \mn@doi [Astronomy {\&} Astrophysics] {10.1051/0004-6361/202244164}, 665, A156

\bibitem[\protect\citeauthoryear{Krasnoperov, Chesnokov, Stark  \& Ravishankara}{Krasnoperov et~al.}{2005}]{Krasnoperov2005}
Krasnoperov L.,  Chesnokov E.,  Stark H.,   Ravishankara A.,  2005, \mn@doi [Proceedings of the Combustion Institute] {10.1016/j.proci.2004.08.223}, 30, 935

\bibitem[\protect\citeauthoryear{Krasnopolsky}{Krasnopolsky}{2012}]{Krasnopolsky2012}
Krasnopolsky V.~A.,  2012, \mn@doi [Icarus] {10.1016/j.icarus.2011.11.012}, 218, 230

\bibitem[\protect\citeauthoryear{Krissansen-Totton, Olson  \& Catling}{Krissansen-Totton et~al.}{2018a}]{Krissansen-Totton2018earth-biosig}
Krissansen-Totton J.,  Olson S.,   Catling D.~C.,  2018a, \mn@doi [Science Advances] {10.1126/sciadv.aao5747}, 4

\bibitem[\protect\citeauthoryear{Krissansen-Totton, Garland, Irwin  \& Catling}{Krissansen-Totton et~al.}{2018b}]{Krissansen-Totton2018t1e-biosig}
Krissansen-Totton J.,  Garland R.,  Irwin P.,   Catling D.~C.,  2018b, \mn@doi [The Astronomical Journal] {10.3847/1538-3881/aad564}, 156, 114

\bibitem[\protect\citeauthoryear{Kurbanov \& Mamedov}{Kurbanov \& Mamedov}{1995}]{KurbanovMamedov1995}
Kurbanov M.~A.,  Mamedov K.~F.,  1995, Kinetics and Catalysis, 36

\bibitem[\protect\citeauthoryear{Langford \& Oldershaw}{Langford \& Oldershaw}{1972}]{LangfordOldershaw1972}
Langford R.~B.,  Oldershaw G.~A.,  1972, \mn@doi [Journal of the Chemical Society, Faraday Transactions 1: Physical Chemistry in Condensed Phases] {10.1039/f19726801550}, 68, 1550

\bibitem[\protect\citeauthoryear{Lee, Stief  \& Timmons}{Lee et~al.}{1977}]{Lee1977}
Lee J.~H.,  Stief L.~J.,   Timmons R.~B.,  1977, \mn@doi [The Journal of Chemical Physics] {10.1063/1.435005}, 67, 1705

\bibitem[\protect\citeauthoryear{Lenton \& Watson}{Lenton \& Watson}{2011}]{LentonWatson2011}
Lenton T.,  Watson A.,  2011, {Revolutions that made the Earth}.
Oxford University Press, \mn@doi{10.1093/acprof:oso/9780199587049.001.0001}

\bibitem[\protect\citeauthoryear{Lessner et~al.,}{Lessner et~al.}{2006}]{Lessner2006}
Lessner D.~J.,  et~al., 2006, \mn@doi [Proceedings of the National Academy of Sciences] {10.1073/pnas.0608833103}, 103, 17921

\bibitem[\protect\citeauthoryear{Lichtin, Berman  \& Lin}{Lichtin et~al.}{1984}]{Lichtin1984}
Lichtin D.,  Berman M.,   Lin M.,  1984, \mn@doi [Chemical Physics Letters] {10.1016/0009-2614(84)80360-7}, 108, 18

\bibitem[\protect\citeauthoryear{Lincowski, Meadows, Crisp, Robinson, Luger, Lustig-Yaeger  \& Arney}{Lincowski et~al.}{2018}]{Lincowski2018}
Lincowski A.~P.,  Meadows V.~S.,  Crisp D.,  Robinson T.~D.,  Luger R.,  Lustig-Yaeger J.,   Arney G.~N.,  2018, \mn@doi [The Astrophysical Journal] {10.3847/1538-4357/aae36a}, 867, 76

\bibitem[\protect\citeauthoryear{Lincowski et~al.,}{Lincowski et~al.}{2023}]{Lincowski2023}
Lincowski A.~P.,  et~al., 2023, The Astrophysical Journal Letters

\bibitem[\protect\citeauthoryear{Liss \& Slater}{Liss \& Slater}{1974}]{LissSlater1974}
Liss P.~S.,  Slater P.~G.,  1974, \mn@doi [Nature] {10.1038/247181a0}, 247, 181

\bibitem[\protect\citeauthoryear{Loison, Halvick, Bergeat, Hickson  \& Wakelam}{Loison et~al.}{2012}]{Loison2012}
Loison J.-C.,  Halvick P.,  Bergeat A.,  Hickson K.~M.,   Wakelam V.,  2012, \mn@doi [Monthly Notices of the Royal Astronomical Society] {10.1111/j.1365-2966.2012.20412.x}, 421, 1476

\bibitem[\protect\citeauthoryear{Louren{\c{c}}o \& Rozel}{Louren{\c{c}}o \& Rozel}{2023}]{LourencoRozel2023}
Louren{\c{c}}o D.~L.,  Rozel A.~B.,  2023, in , Dynamics of Plate Tectonics and Mantle Convection.
Elsevier, pp 181--196, \mn@doi{10.1016/B978-0-323-85733-8.00004-4}

\bibitem[\protect\citeauthoryear{Louren{\c{c}}o, Rozel, Ballmer  \& Tackley}{Louren{\c{c}}o et~al.}{2020}]{Lourenco2020}
Louren{\c{c}}o D.~L.,  Rozel A.~B.,  Ballmer M.~D.,   Tackley P.~J.,  2020, \mn@doi [Geochemistry, Geophysics, Geosystems] {10.1029/2019GC008756}, 21

\bibitem[\protect\citeauthoryear{Lu, Wu, Lee, Zhu  \& Lin}{Lu et~al.}{2006}]{Lu2006}
Lu C.-W.,  Wu Y.-J.,  Lee Y.-P.,  Zhu R.~S.,   Lin M.~C.,  2006, \mn@doi [The Journal of Chemical Physics] {10.1063/1.2357739}, 125

\bibitem[\protect\citeauthoryear{Luger \& Barnes}{Luger \& Barnes}{2015}]{LugerBarnes2015}
Luger R.,  Barnes R.,  2015, \mn@doi [Astrobiology] {10.1089/ast.2014.1231}, 15, 119

\bibitem[\protect\citeauthoryear{Mak et~al.,}{Mak et~al.}{2024}]{Mak2024}
Mak M.~T.,  et~al., 2024, Submitted to MNRAS

\bibitem[\protect\citeauthoryear{Manabe \& Strickler}{Manabe \& Strickler}{1964}]{ManabeStrickler1964}
Manabe S.,  Strickler R.~F.,  1964, \mn@doi [Journal of the Atmospheric Sciences] {10.1175/1520-0469(1964)021<0361:TEOTAW>2.0.CO;2}, 21, 361

\bibitem[\protect\citeauthoryear{Manabe \& Wetherald}{Manabe \& Wetherald}{1967}]{ManabeWetherald1967}
Manabe S.,  Wetherald R.~T.,  1967, \mn@doi [Journal of the Atmospheric Sciences] {10.1175/1520-0469(1967)024<0241:TEOTAW>2.0.CO;2}, 24, 241

\bibitem[\protect\citeauthoryear{Manners, Edwards, Hill  \& Thelen}{Manners et~al.}{2022}]{Manners2022}
Manners J.,  Edwards J.~M.,  Hill P.,   Thelen J.-C.,  2022, {SOCRATES (Suite Of Community RAdiative Transfer Codes Based on Edwards and Slingo) Technical Guide}

\bibitem[\protect\citeauthoryear{Martinez \& Herron}{Martinez \& Herron}{1983}]{MartinezHerron1983}
Martinez R.~I.,  Herron J.~T.,  1983, \mn@doi [International Journal of Chemical Kinetics] {10.1002/kin.550151102}, 15, 1127

\bibitem[\protect\citeauthoryear{McCollom \& Bach}{McCollom \& Bach}{2009}]{McCollomBach2009}
McCollom T.~M.,  Bach W.,  2009, \mn@doi [Geochimica et Cosmochimica Acta] {10.1016/j.gca.2008.10.032}, 73, 856

\bibitem[\protect\citeauthoryear{Meadows et~al.,}{Meadows et~al.}{2018}]{Meadows2018pc}
Meadows V.~S.,  et~al., 2018, \mn@doi [Astrobiology] {10.1089/ast.2016.1589}, 18, 133

\bibitem[\protect\citeauthoryear{Morii, Koga  \& Nagai}{Morii et~al.}{1987}]{Morii1987}
Morii H.,  Koga Y.,   Nagai S.,  1987, \mn@doi [Biotechnology and Bioengineering] {10.1002/bit.260290304}, 29, 310

\bibitem[\protect\citeauthoryear{Nicholas, Amodio  \& Baker}{Nicholas et~al.}{1979}]{Nicholas1979}
Nicholas J.~E.,  Amodio C.~A.,   Baker M.~J.,  1979, \mn@doi [Journal of the Chemical Society, Faraday Transactions 1: Physical Chemistry in Condensed Phases] {10.1039/f19797501868}, 75, 1868

\bibitem[\protect\citeauthoryear{Nicholson, Daines, Mayne, Eager-Nash, Lenton  \& Kohary}{Nicholson et~al.}{2022}]{Nicholson2022}
Nicholson A.~E.,  Daines S.~J.,  Mayne N.~J.,  Eager-Nash J.~K.,  Lenton T.~M.,   Kohary K.,  2022, \mn@doi [Monthly Notices of the Royal Astronomical Society] {10.1093/mnras/stac2086}, 517, 222

\bibitem[\protect\citeauthoryear{Olson, Jansen  \& Abbot}{Olson et~al.}{2020}]{Olson2020}
Olson S.~L.,  Jansen M.,   Abbot D.~S.,  2020, \mn@doi [The Astrophysical Journal] {10.3847/1538-4357/ab88c9}, 895, 19

\bibitem[\protect\citeauthoryear{Ozaki, Tajika, Hong, Nakagawa  \& Reinhard}{Ozaki et~al.}{2018}]{Ozaki2018}
Ozaki K.,  Tajika E.,  Hong P.~K.,  Nakagawa Y.,   Reinhard C.~T.,  2018, \mn@doi [Nature Geoscience] {10.1038/s41561-017-0031-2}, 11, 55

\bibitem[\protect\citeauthoryear{Palin \& Santosh}{Palin \& Santosh}{2021}]{PalinSantosh2021}
Palin R.~M.,  Santosh M.,  2021, \mn@doi [Gondwana Research] {10.1016/j.gr.2020.11.001}, 100, 3

\bibitem[\protect\citeauthoryear{Peacock, Barman, Shkolnik, Hauschildt  \& Baron}{Peacock et~al.}{2019}]{Peacock2019}
Peacock S.,  Barman T.,  Shkolnik E.~L.,  Hauschildt P.~H.,   Baron E.,  2019, \mn@doi [The Astrophysical Journal] {10.3847/1538-4357/aaf891}, 871, 235

\bibitem[\protect\citeauthoryear{Peng, Hu  \& Marshall}{Peng et~al.}{1999}]{Peng1999}
Peng J.,  Hu X.,   Marshall P.,  1999, \mn@doi [The Journal of Physical Chemistry A] {10.1021/jp984242l}, 103, 5307

\bibitem[\protect\citeauthoryear{Pinto, Gladstone  \& Yung}{Pinto et~al.}{1980}]{Pinto1980}
Pinto J.~P.,  Gladstone G.~R.,   Yung Y.~L.,  1980, \mn@doi [Science] {10.1126/science.210.4466.183}, 210, 183

\bibitem[\protect\citeauthoryear{Ranjan, Schwieterman, Harman, Fateev, Sousa-Silva, Seager  \& Hu}{Ranjan et~al.}{2020}]{Ranjan2020}
Ranjan S.,  Schwieterman E.~W.,  Harman C.,  Fateev A.,  Sousa-Silva C.,  Seager S.,   Hu R.,  2020, \mn@doi [The Astrophysical Journal] {10.3847/1538-4357/ab9363}, 896, 148

\bibitem[\protect\citeauthoryear{Ranjan, Schwieterman, Leung, Harman  \& Hu}{Ranjan et~al.}{2023}]{Ranjan2023}
Ranjan S.,  Schwieterman E.~W.,  Leung M.,  Harman C.~E.,   Hu R.,  2023, Submitted to AAS Journals

\bibitem[\protect\citeauthoryear{Ridgway et~al.,}{Ridgway et~al.}{2022}]{Ridgway2022}
Ridgway R.~J.,  et~al., 2022, \mn@doi [Monthly Notices of the Royal Astronomical Society] {10.1093/mnras/stac3105}, 518, 2472

\bibitem[\protect\citeauthoryear{Rosing}{Rosing}{1999}]{Rosing1999}
Rosing M.~T.,  1999, \mn@doi [Science] {10.1126/science.283.5402.674}, 283, 674

\bibitem[\protect\citeauthoryear{Rothman et~al.,}{Rothman et~al.}{2013}]{Rothman2013}
Rothman L.~S.,  et~al., 2013, \mn@doi [Journal of Quantitative Spectroscopy and Radiative Transfer] {https://doi.org/10.1016/j.jqsrt.2013.07.002}, 130, 4

\bibitem[\protect\citeauthoryear{Rugheimer \& Kaltenegger}{Rugheimer \& Kaltenegger}{2018}]{Rugheimer2018}
Rugheimer S.,  Kaltenegger L.,  2018, \mn@doi [The Astrophysical Journal] {10.3847/1538-4357/aaa47a}, 854, 19

\bibitem[\protect\citeauthoryear{Rugheimer, Kaltenegger, Segura, Linsky  \& Mohanty}{Rugheimer et~al.}{2015}]{Rugheimer2015}
Rugheimer S.,  Kaltenegger L.,  Segura A.,  Linsky J.,   Mohanty S.,  2015, \mn@doi [The Astrophysical Journal] {10.1088/0004-637X/809/1/57}, 809, 57

\bibitem[\protect\citeauthoryear{Salazar, Olson, Komacek, Stephens  \& Abbot}{Salazar et~al.}{2020}]{Salazar2020}
Salazar A.~M.,  Olson S.~L.,  Komacek T.~D.,  Stephens H.,   Abbot D.~S.,  2020, \mn@doi [The Astrophysical Journal] {10.3847/2041-8213/ab94c1}, 896

\bibitem[\protect\citeauthoryear{Sauterey, Charnay, Affholder, Mazevet  \& Ferri{\`{e}}re}{Sauterey et~al.}{2020}]{Sauterey2020}
Sauterey B.,  Charnay B.,  Affholder A.,  Mazevet S.,   Ferri{\`{e}}re R.,  2020, \mn@doi [Nature Communications] {10.1038/s41467-020-16374-7}, 11, 2705

\bibitem[\protect\citeauthoryear{Schofield}{Schofield}{1973}]{Schofield1973}
Schofield K.,  1973, \mn@doi [Journal of Physical and Chemical Reference Data] {10.1063/1.3253112}, 2, 25

\bibitem[\protect\citeauthoryear{Schonheit, Moll  \& Thauer}{Schonheit et~al.}{1980}]{Schonheit1980}
Schonheit P.,  Moll J.,   Thauer R.~K.,  1980, \mn@doi [Archives of Microbiology] {10.1007/BF00414356}, 127, 59

\bibitem[\protect\citeauthoryear{Schwieterman, Reinhard, Olson, Ozaki, Harman, Hong  \& Lyons}{Schwieterman et~al.}{2019}]{Schwieterman2019}
Schwieterman E.~W.,  Reinhard C.~T.,  Olson S.~L.,  Ozaki K.,  Harman C.~E.,  Hong P.~K.,   Lyons T.~W.,  2019, \mn@doi [The Astrophysical Journal] {10.3847/1538-4357/ab05e1}, 874, 9

\bibitem[\protect\citeauthoryear{Segura, Kasting, Meadows, Cohen, Scalo, Crisp, Butler  \& Tinetti}{Segura et~al.}{2005}]{Segura2005}
Segura A.,  Kasting J.~F.,  Meadows V.,  Cohen M.,  Scalo J.,  Crisp D.,  Butler R.~A.,   Tinetti G.,  2005, \mn@doi [Astrobiology] {10.1089/ast.2005.5.706}, 5, 706

\bibitem[\protect\citeauthoryear{Shields, Meadows, Bitz, Pierrehumbert, Joshi  \& Robinson}{Shields et~al.}{2013}]{Shields2013}
Shields A.~L.,  Meadows V.~S.,  Bitz C.~M.,  Pierrehumbert R.~T.,  Joshi M.~M.,   Robinson T.~D.,  2013, \mn@doi [Astrobiology] {10.1089/ast.2012.0961}, 13, 715

\bibitem[\protect\citeauthoryear{Singleton \& Cvetanovi{\'{c}}}{Singleton \& Cvetanovi{\'{c}}}{1988}]{SingletonCvetanovic1988}
Singleton D.~L.,  Cvetanovi{\'{c}} R.~J.,  1988, \mn@doi [Journal of Physical and Chemical Reference Data] {10.1063/1.555811}, 17, 1377

\bibitem[\protect\citeauthoryear{Slanger, Wood  \& Black}{Slanger et~al.}{1972}]{Slanger1972}
Slanger T.~G.,  Wood B.~J.,   Black G.,  1972, \mn@doi [The Journal of Chemical Physics] {10.1063/1.1677953}, 57, 233

\bibitem[\protect\citeauthoryear{Solomatov}{Solomatov}{1995}]{Solomatov1995}
Solomatov V.~S.,  1995, \mn@doi [Physics of Fluids] {10.1063/1.868624}, 7, 266

\bibitem[\protect\citeauthoryear{Stachnik \& Molina}{Stachnik \& Molina}{1987}]{StachnikMolina1897}
Stachnik R.~A.,  Molina M.~J.,  1987, \mn@doi [The Journal of Physical Chemistry] {10.1021/j100301a035}, 91, 4603

\bibitem[\protect\citeauthoryear{Teal, Kempton, Bastelberger, Youngblood  \& Arney}{Teal et~al.}{2022}]{Teal2022}
Teal D.~J.,  Kempton E. M.-R.,  Bastelberger S.,  Youngblood A.,   Arney G.,  2022, \mn@doi [The Astrophysical Journal] {10.3847/1538-4357/ac4d99}, 927, 90

\bibitem[\protect\citeauthoryear{Thiesemann, MacNamara  \& Taatjes}{Thiesemann et~al.}{1997}]{Thiesemann1997}
Thiesemann H.,  MacNamara J.,   Taatjes C.~A.,  1997, \mn@doi [The Journal of Physical Chemistry A] {10.1021/jp9630333}, 101, 1881

\bibitem[\protect\citeauthoryear{Thompson, Krissansen-Totton, Wogan, Telus  \& Fortney}{Thompson et~al.}{2022}]{Thompson2022}
Thompson M.~A.,  Krissansen-Totton J.,  Wogan N.,  Telus M.,   Fortney J.~J.,  2022, \mn@doi [Proceedings of the National Academy of Sciences] {10.1073/pnas.2117933119}, 119

\bibitem[\protect\citeauthoryear{Tian, France, Linsky, Mauas  \& Vieytes}{Tian et~al.}{2014}]{Tian2014}
Tian F.,  France K.,  Linsky J.~L.,  Mauas P.~J.,   Vieytes M.~C.,  2014, \mn@doi [Earth and Planetary Science Letters] {10.1016/j.epsl.2013.10.024}, 385, 22

\bibitem[\protect\citeauthoryear{Tiee, Wampler, Oldenborg  \& Rice}{Tiee et~al.}{1981}]{Tiee1981}
Tiee J.,  Wampler F.,  Oldenborg R.,   Rice W.,  1981, \mn@doi [Chemical Physics Letters] {10.1016/0009-2614(81)85111-1}, 82, 80

\bibitem[\protect\citeauthoryear{Toon, Kasting, Turco  \& Liu}{Toon et~al.}{1987}]{Toon1987}
Toon O.~B.,  Kasting J.~F.,  Turco R.~P.,   Liu M.~S.,  1987, \mn@doi [Journal of Geophysical Research] {10.1029/JD092iD01p00943}, 92, 943

\bibitem[\protect\citeauthoryear{Tsang \& Hampson}{Tsang \& Hampson}{1986}]{TsangHampson1986}
Tsang W.,  Hampson R.~F.,  1986, \mn@doi [Journal of Physical and Chemical Reference Data] {10.1063/1.555759}, 15, 1087

\bibitem[\protect\citeauthoryear{Tsang \& Herron}{Tsang \& Herron}{1991}]{TsangHerron1991}
Tsang W.,  Herron J.~T.,  1991, \mn@doi [Journal of Physical and Chemical Reference Data] {10.1063/1.555890}, 20, 609

\bibitem[\protect\citeauthoryear{Turbet, Boulet  \& Karman}{Turbet et~al.}{2020}]{Turbet2020}
Turbet M.,  Boulet C.,   Karman T.,  2020, \mn@doi [Icarus] {https://doi.org/10.1016/j.icarus.2020.113762}, 346, 113762

\bibitem[\protect\citeauthoryear{Turco, Whitten  \& Toon}{Turco et~al.}{1982}]{Turco1982}
Turco R.~P.,  Whitten R.~C.,   Toon O.~B.,  1982, \mn@doi [Reviews of Geophysics] {10.1029/RG020i002p00233}, 20, 233

\bibitem[\protect\citeauthoryear{Villanueva, Smith, Protopapa, Faggi  \& Mandell}{Villanueva et~al.}{2018}]{Villanueva2018}
Villanueva G.,  Smith M.,  Protopapa S.,  Faggi S.,   Mandell A.,  2018, \mn@doi [Journal of Quantitative Spectroscopy and Radiative Transfer] {10.1016/j.jqsrt.2018.05.023}, 217, 86

\bibitem[\protect\citeauthoryear{Walker, Hays  \& Kasting}{Walker et~al.}{1981}]{Walker1981}
Walker J. C.~G.,  Hays P.~B.,   Kasting J.~F.,  1981, \mn@doi [Journal of Geophysical Research] {10.1029/JC086iC10p09776}, 86, 9776

\bibitem[\protect\citeauthoryear{Ward, Rasmussen  \& Fischer}{Ward et~al.}{2019}]{Ward2019}
Ward L.~M.,  Rasmussen B.,   Fischer W.~W.,  2019, \mn@doi [Journal of Geophysical Research: Biogeosciences] {10.1029/2018JG004679}, 124, 211

\bibitem[\protect\citeauthoryear{Watson}{Watson}{2008}]{Watson2008}
Watson A.~J.,  2008, \mn@doi [Astrobiology] {10.1089/ast.2006.0115}, 8, 175

\bibitem[\protect\citeauthoryear{Weiss, Sousa, Mrnjavac, Neukirchen, Roettger, Nelson-Sathi  \& Martin}{Weiss et~al.}{2016}]{Weiss2016}
Weiss M.~C.,  Sousa F.~L.,  Mrnjavac N.,  Neukirchen S.,  Roettger M.,  Nelson-Sathi S.,   Martin W.~F.,  2016, \mn@doi [Nature Microbiology] {10.1038/nmicrobiol.2016.116}, 1, 16116

\bibitem[\protect\citeauthoryear{Wilson et~al.,}{Wilson et~al.}{2021}]{Wilson2021}
Wilson D.~J.,  et~al., 2021, \mn@doi [The Astrophysical Journal] {10.3847/1538-4357/abe771}, 911, 18

\bibitem[\protect\citeauthoryear{Wine, Chameides  \& Ravishankara}{Wine et~al.}{1981}]{Wine1981}
Wine P.~H.,  Chameides W.~L.,   Ravishankara A.~R.,  1981, \mn@doi [Geophysical Research Letters] {10.1029/GL008i005p00543}, 8, 543

\bibitem[\protect\citeauthoryear{Wogan, Catling, Zahnle  \& Claire}{Wogan et~al.}{2022}]{Wogan2022}
Wogan N.~F.,  Catling D.~C.,  Zahnle K.~J.,   Claire M.~W.,  2022, \mn@doi [Proceedings of the National Academy of Sciences] {10.1073/pnas.2205618119}, 119

\bibitem[\protect\citeauthoryear{Wunderlich et~al.,}{Wunderlich et~al.}{2020}]{Wunderlich2020}
Wunderlich F.,  et~al., 2020, \mn@doi [The Astrophysical Journal] {10.3847/1538-4357/aba59c}, 901, 126

\bibitem[\protect\citeauthoryear{Yates, Palmer, Manners, Boutle, Kohary, Mayne  \& Abraham}{Yates et~al.}{2020}]{Yates2020}
Yates J.~S.,  Palmer P.~I.,  Manners J.,  Boutle I.,  Kohary K.,  Mayne N.,   Abraham L.,  2020, \mn@doi [Monthly Notices of the Royal Astronomical Society] {10.1093/mnras/stz3520}, 492

\bibitem[\protect\citeauthoryear{Yung \& Demore}{Yung \& Demore}{1982}]{YungDemore1982}
Yung Y.~L.,  Demore W.,  1982, \mn@doi [Icarus] {10.1016/0019-1035(82)90080-X}, 51, 199

\bibitem[\protect\citeauthoryear{Yung, Allen  \& Pinto}{Yung et~al.}{1984}]{Yung1984}
Yung Y.~L.,  Allen M.,   Pinto J.~P.,  1984, \mn@doi [The Astrophysical Journal Supplement Series] {10.1086/190963}, 55, 465

\bibitem[\protect\citeauthoryear{Zabarnick, Fleming  \& Lin}{Zabarnick et~al.}{1989}]{Zabarnick1989}
Zabarnick S.,  Fleming J.~W.,   Lin M.~C.,  1989, \mn@doi [International Journal of Chemical Kinetics] {10.1002/kin.550210905}, 21, 765

\bibitem[\protect\citeauthoryear{Zahnle}{Zahnle}{1986}]{Zahnle1986}
Zahnle K.~J.,  1986, \mn@doi [Journal of Geophysical Research: Atmospheres] {https://doi.org/10.1029/JD091iD02p02819}, 91, 2819

\bibitem[\protect\citeauthoryear{Zahnle, Claire  \& Catling}{Zahnle et~al.}{2006}]{Zahnle2006}
Zahnle K.,  Claire M.,   Catling D.,  2006, \mn@doi [Geobiology] {10.1111/j.1472-4669.2006.00085.x}, 4, 271

\bibitem[\protect\citeauthoryear{Zieba et~al.,}{Zieba et~al.}{2023}]{Zieba2023}
Zieba S.,  et~al., 2023, \mn@doi [Nature] {10.1038/s41586-023-06232-z}

\makeatother
\end{thebibliography}
